\documentclass[11pt]{article}
\usepackage{enumitem}
\usepackage{lineno}

\usepackage[utf8x]{inputenc}
\usepackage[T1]{fontenc}
\usepackage[english]{babel}
\usepackage{csquotes}
\usepackage[
            sorting=ync, 
            backend=biber, 
            style=authoryear-comp, 
            uniquelist=false,
            uniquename=false,
            natbib,
            maxcitenames=2,
            maxbibnames=99]{biblatex}

\renewbibmacro{in:}{%
  \ifentrytype{article}{}{\printtext{\bibstring{in}\intitlepunct}}}

\DeclareNameFormat{always-init}{%
	\ifnumequal{\value{uniquename}}{2}
	{\usebibmacro{name:family-given}
		{\namepartfamily}
		{\namepartgiven}
		{\namepartprefix}
		{\namepartsuffix}}
	{\usebibmacro{name:family-given}
		{\namepartfamily}
		{\namepartgiveni}
		{\namepartprefix}
		{\namepartsuffix}}%
	\usebibmacro{name:andothers}}

\DeclareNameFormat{family_only}{%
  \usebibmacro{name:family}
    {\namepartfamily}
    {\namepartgiven}
    {\namepartprefix}
    {\namepartsuffix}%
  \usebibmacro{name:andothers}}

\DeclareNameAlias{author}{always-init}
\DeclareNameAlias{editor}{always-init}

\DeclareSortingNamekeyTemplate{
	\keypart{\namepart{family}}
	\keypart{\namepart{prefix}}
	\keypart{\namepart{given}}
	\keypart{\namepart{suffix}}}

\DeclareSortingTemplate{nyc}{
  \sort{
    \field{presort}
  }
  \sort[final]{
    \field{sortkey}
  }
  \sort{
    \field{sortname}
    \field{author}
    \field{editor}
    \field{translator}
    \field{sorttitle}
    \field{title}
  }
  \sort{
    \field{sortyear}
    \field{year}
  }
  \sort{\citeorder}
}

\DeclareSortingTemplate{ync}{
  \sort{
    \field{presort}
  }
  \sort[final]{
    \field{sortkey}
  }
   \sort{
    \field{sortyear}
    \field{year}
  }
  \sort{
    \field{sortname}
    \field{author}
    \field{editor}
    \field{translator}
    \field{sorttitle}
    \field{title}
  }
  \sort{\citeorder}
}

\AtEveryBibitem{\clearfield{month}}
\AtEveryBibitem{\clearfield{day}}
\AtEveryBibitem{\clearfield{url}}
\AtEveryBibitem{\clearfield{urlyear}}
\AtEveryBibitem{\clearfield{issn}}

\DeclareNameWrapperFormat{labelname:poss}{#1's}

\newrobustcmd*{\posscitealias}{%
	\AtNextCite{%
		\DeclareNameWrapperAlias{labelname}{labelname:poss}}}

\newrobustcmd*{\posscite}{%
	\posscitealias
	\textcite}

\newrobustcmd*{\Posscite}{\bibsentence\posscite}

 \DeclareNameWrapperFormat{labelname:posss}{#1'}

\newrobustcmd*{\possscitealias}{%
	\AtNextCite{%
		\DeclareNameWrapperAlias{labelname}{labelname:posss}}}

\newrobustcmd*{\possscite}{%
	\possscitesalias
	\textcite}

\newrobustcmd*{\Possscite}{\bibsentence\possscites}

\newrobustcmd*{\posscites}{%
	\possscitealias
	\textcites}

\usepackage[dvipsnames,table,xcdraw]{xcolor} 

\usepackage{ragged2e} 

\usepackage{amsthm, amsmath, amssymb} 
\usepackage{mathrsfs} 
\usepackage{float} 
\usepackage{newfloat} 
\usepackage{graphicx, multicol} 
\usepackage{framed}
\usepackage[normalem]{ulem} 

\usepackage[labelfont=bf]{caption} 

\usepackage{amsfonts}
\usepackage{enumitem}
\usepackage{mathtools}
\usepackage{multicol}
\usepackage{color,soul}
\usepackage{relsize} 
\usepackage{fancyhdr}
\usepackage{titletoc}
\usepackage{afterpage}
\usepackage{bigstrut}

\usepackage{makecell,tabularx}

\usepackage{rotating}
\usepackage{setspace}

\usepackage{lipsum}

\usepackage{tikz}

\newcounter{cases}
\newcounter{subcases}[cases]

\newcommand{\myfig}[4]{\begin{figure}[H]\centering \begin{center} \includegraphics[width=#1\textwidth]{#2} \caption{#3} \label{#4} \end{center} \end{figure}}

\usepackage[most]{tcolorbox}

\newtcolorbox[auto counter,]{fancybox}[3][]{
arc=5mm,
lower separated=false,
fonttitle=\bfseries,
colbacktitle=black!10,
coltitle=black,
colupper=black,
enhanced,
attach boxed title to top left={xshift=3cm,
        yshift=-3mm},
colframe=black,
colback=black!10,
title=#2 \thetcbcounter : #3,#1,breakable}

\newcounter{romanequation}

\newlength{\depthofsumsign}
\usepackage{tokcycle}[2021-03-10]
\usepackage{xcolor}
\newcounter{wordcount}
\newcounter{lettercount}
\newcounter{wordlimit}
\newif\ifinword
\newif\ifrunningcount
\newif\ifsummarycount
\def\limitcolor{red}
\makeatletter
\newcommand\addtomacro[2]{\tc@defx#1{#1#2}}
\newcommand\changecolor[1]{\tctestifx{.#1}{}{\addcytoks{\color{#1}{}}%
  \tc@defx\currentcolor{#1}}}
\makeatother
\newcommand\dumpword{%
  \addcytoks[1]{\accumword}%
  \ifinword\stepcounter{wordcount}
    \ifrunningcount\addcytoks[x]{$^{\thewordcount,\thelettercount}$}\fi
    \ifnum\thewordcount=\value{wordlimit}\relax\changecolor{\limitcolor}\fi
  \fi%
  \inwordfalse
  \def\accumword{}}
\newcommand\addletter[1]{%
  \tctestifcatnx A#1{\stepcounter{lettercount}\inwordtrue}{\dumpword}%
  \addtomacro\accumword{#1}}
\xtokcycleenvironment\countem
  {\addletter{##1}}
  {\dumpword\groupedcytoks{\processtoks{##1}\dumpword\expandafter}\expandafter
    \changecolor\expandafter{\currentcolor}}
  {\dumpword\addcytoks{##1}}
  {\dumpword\addcytoks{##1}}
  {\stripgroupingtrue\def\accumword{}\def\currentcolor{.}
    \setcounter{wordcount}{0}\setcounter{lettercount}{0}}
  {\dumpword\ifsummarycount\tcafterenv{%
    \par(Wordcount=\thewordcount, Lettercount=\thelettercount)}\fi}
    
\usepackage{xr-hyper}

\makeatletter
\newcommand*{\addFileDependency}[1]{
\typeout{(#1)}
\@addtofilelist{#1}
\IfFileExists{#1}{}{\typeout{No file #1.}}
}\makeatother

\usepackage{thmtools}
\usepackage[framemethod=TikZ]{mdframed}
\usepackage{tcolorbox}

\theoremstyle{definition}
\mdfdefinestyle{mdbluebox}{%
	roundcorner = 10pt,
	linewidth=1pt,
	skipabove=12pt,
	innerbottommargin=9pt,
	skipbelow=2pt,
	nobreak=true,
	linecolor=blue,
	backgroundcolor=Blue!5,
}
\declaretheoremstyle[
	headfont=\sffamily\bfseries\color{MidnightBlue},
	mdframed={style=mdbluebox},
	headpunct={\\[3pt]},
	postheadspace={0pt}
]{thmbluebox}

\mdfdefinestyle{mdredbox}{%
	linewidth=0.5pt,
	skipabove=12pt,
	frametitleaboveskip=5pt,
	frametitlebelowskip=0pt,
	skipbelow=2pt,
	frametitlefont=\bfseries,
	innertopmargin=4pt,
	innerbottommargin=8pt,
	linecolor=RawSienna,
	backgroundcolor=OrangeRed!5,
}

\declaretheoremstyle[
	headfont=\bfseries\color{RawSienna},
	mdframed={style=mdredbox},
	headpunct={\\[3pt]},
	postheadspace={0pt},
]{thmredbox}

\declaretheoremstyle[
	headfont=\bfseries\color{BurntOrange!50!black},
	mdframed={style=mdyellowbox},
	headpunct={\\[3pt]},
	postheadspace={0pt},
]{thmyellowbox}

\mdfdefinestyle{mdyellowbox}{%
	linewidth=0.5pt,
	skipabove=12pt,
	frametitleaboveskip=5pt,
	frametitlebelowskip=0pt,
	skipbelow=2pt,
	frametitlefont=\bfseries,
	innertopmargin=4pt,
	innerbottommargin=8pt,
	linecolor=BurntOrange,
	backgroundcolor=Goldenrod!5,
}

\usepackage{nameref} 

\mdfdefinestyle{mdgreenbox}{%
	linewidth=0.5pt,
	skipabove=12pt,
	frametitleaboveskip=5pt,
	frametitlebelowskip=0pt,
	skipbelow=2pt,
	frametitlefont=\bfseries,
	innertopmargin=4pt,
	innerbottommargin=8pt,
	linecolor=ForestGreen,
	backgroundcolor=ForestGreen!5,
}

\declaretheoremstyle[%
	headfont=\bfseries\color{ForestGreen!70!black},
	mdframed={style=mdgreenbox},
	headpunct={\\[3pt]},
	postheadspace={0pt},
    headformat=\NOTE,
    notebraces={}{},
    notefont = \bfseries\color{ForestGreen!70!black}
]{thmgreenbox}

\makeatletter
\newcommand{\myitem}[1]{%
\item[#1]\protected@edef\@currentlabel{#1}%
}
\makeatother
\usepackage{empheq}

\usepackage[final, colorlinks = true, 
            linkcolor = ForestGreen, 
            citecolor = MidnightBlue,
            filecolor = black,
            hypertexnames = false,
            urlcolor = MidnightBlue,
            breaklinks=true,
            pdfencoding=auto, psdextra]{hyperref} 

\usepackage{fullpage}
\usepackage[title]{appendix} 
\usepackage{etoolbox, apptools}
\AtBeginEnvironment{appendices}{\appendixtrue}

\usepackage{titlesec}
\usepackage{titletoc}
\titleformat{\section}[block]{\Large\bfseries\filcenter}{\IfAppendix{\appendixname~}{\relax}\thesection\IfAppendix{: }{}}{1em}{}
\titleformat{\subsection}[block]{\Large\itshape\filcenter}{\thesubsection}{1em}{}
\titleformat{\subsubsection}[block]{\large\itshape}{\thesubsubsection}{1em}{}
\titleformat{\paragraph}[runin]{\itshape}{\theparagraph}{1em}{}[. ]

\usepackage{fancyhdr}

\renewenvironment{abstract}
{\small
\begin{center}
	\bfseries \abstractname\vspace{-.5em}\vspace{0pt}
\end{center}
\list{}{
	\setlength{\leftmargin}{0cm}
	\setlength{\rightmargin}{\leftmargin}%
}%
\item\relax}
{\endlist}

\title{Demographic senescence as multi-level selection in miniature
 }

\author{ %
	Ananda Shikhara Bhat$^{1,2,\ast}$ \and Hanna Kokko$^{1,2,\dag}$
}

\date{
\small
1. Institute of Organismic and Molecular Evolution (iomE), Johannes Gutenberg University, 55128 Mainz, Germany; 2. Institute for Quantitative and Computational Biosciences (IQCB), Johannes Gutenberg University, 55128 Mainz, Germany\\
$\ast$ E-mail: abhat@uni-mainz.de \ \ \ \ \ $\dag$ E-mail: hkokko@uni-mainz.de
}

\titlecontents{section}
[0pt]                                               
{}%
{\contentsmargin{0pt}                               
    \thecontentslabel\enspace%
    \large}
{\contentsmargin{0pt}\large}                        
{\titlerule*[.5pc]{.}\contentspage}                 
[] 

\usepackage{xspace}
\newcommand{\lows}{sub-systems\xspace}
\newcommand{\highs}{systems\xspace}
\newcommand{\low}{sub-system\xspace}
\newcommand{\high}{system\xspace}
\newcommand{\fails}{{\color{red}red}\xspace}
\newcommand{\notfails}{{\color{blue}blue}\xspace}

\newcommand{\failure}{{\color{red}warming}\xspace}
\newcommand{\repair}{{\color{blue}cooling}\xspace}

\newcommand{\repprob}{{\varrho}}

\begin{document}


\maketitle 

\begin{abstract}
{\small
\begin{singlespace}


\countem
Multi-level selection and senescence do not at first sight have much in common. Here, we demonstrate that the emergent mortality patterns generated by demographic senescence can be understood as the product of multi-level selection. We formulate a two-level Moran type process and use its scaling limits to illustrate that a simple mathematical framework that models multi-level selection in group-structured populations also models damage accumulation patterns and resultant mortality curves in ageing organisms. To verbally make the connection, observe that defectors spread within a group consisting of cooperators and defectors; when groups compete against each other, defector-rich groups suffer,  and  between-group selection causes such groups to be systematically under-represented. Exactly analogously, senescing individuals accumulate damage to physiological sub-systems, and `damage begets damage'; individuals who are more damaged are more likely to die, hence damage-rich individuals are systematically under-represented in later age classes. Thus, emergent senescence patterns in complex, integrated organisms are formally equivalent to the patterns generated by a within-generation multi-level selection process in which intra-organismal sub-systems play the role of particles, organisms play the role of collectives, and selective disappearance plays the role of group selection.
\endcountem
\end{singlespace}
}
\end{abstract}

{\noindent \small \textbf{Keywords:} altruism; prudence; metapopulations; frailty; reliability theory; selective disappearance}



\begin{refsection}
\section{Introduction}

Imagine a population of hosts and pathogens in which the rate at which the pathogen exploits the host for its own growth is an evolving trait. When pathogens cannot easily be transmitted from one host population to another, they can be selected to become `prudent' even though selection within the host favours a maximally productive phenotype that exploits the host at the maximum possible rate~\citep{kerr_2006_prudent}. Similarly, predators may evolve to be `prudent' in structured landscapes, where those predators that overexploit their prey risk going locally extinct~\citep{mgonigle_2026_prudence}. Genetic elements such as plasmids and early replicators in proto-cells may selfishly replicate faster than the collective organism that they are found in to increase their own fitness, but face a disadvantage when the individual they are contained in is also competing with other individuals~\citep{paulsson_2002_plasmid,cooney_2022_chromosomes,rossine_2025_plasmid}. In game theoretic models of cooperators and defectors organised into groups, defectors outcompete cooperators within a group, but groups comprised of more cooperators are favoured by group-level selection~\citep{boorman_group_1973,kimura_evolution_1984,fontanari_groups_2014,cooney_replicator_2019,cooney_evolutionary_2023}.

All these scenarios consist of a population comprised of two types of entities (`particles' in  multi-level selection parlance), say \fails and \notfails, organised into groups (`collectives' in multi-level selection parlance). One type, say \fails, has a within-group competitive advantage but also increases the vulnerability of the group as a whole to face adverse consequences such as going extinct (in the prudence examples) or being outcompeted (in the game theoretic examples), and thus prominently feature multi-level selection ($\text{MLS}_{2}$ \emph{sensu} \cite{okasha_MLSbook_2006}). Many of the key features of these processes can be captured by a stochastic two-level Moran process implicitly used by \citet{kimura_group_1983,kimura_evolution_1984}, formally introduced by \citet{luo_unifying_2014} and \citet{luo_scaling_2017}, and extended to general two-level two-strategy evolutionary games by \citet{cooney_replicator_2019, cooney_evolutionary_2023} (also see \citet{fontanari_groups_2014,vanveelen_groups_2014,cooney_2022_chromosomes,cooney_2025_cultural} for further extensions and applications to particular multi-level biological systems).


Here, we extend \posscite{luo_unifying_2014} model in a rather different conceptual direction that broadens its biological reach. We construct a two-level stochastic process inspired by \posscite{luo_unifying_2014} framework to show that demographic senescence, the progressive increase in mortality with age, can also be rephrased in the language of \posscite{luo_unifying_2014} multi-level selection models. In particular, we show that models in which individual-level mortality risk, the manifestation of vulnerability in a demographic context, depends on the functioning of various intra-organismal `sub-systems'~\citep{gavrilov_reliability_2001,bhat_failures_2026} or the stochastic variation of an abstract `physiological state'~\citep{woodbury_random-walk_1977,yashin_mortality_1985,weitz_2001_explaining}, can be formally viewed as the outcome of selection operating simultaneously at two levels.


The conceptual analogy between multi-level selection and  demographic senescence (due to failure accumulation) is made by observing that an organism with some fraction of failed/damaged sub-systems tends to experience ever increasing failure risk because intra-organismal sub-systems are often interdependent, and thus, when one sub-system fails, the sub-systems that depend on it become even more likely to fail than before~\citep{gavrilov_biology_1991,ledberg_exponential_2020,nielsen_gompertz_2024,bhat_failures_2026}. At the same time, those organisms with more failed sub-systems also experience higher mortality risk~\citep{gavrilov_biology_1991,bhat_failures_2026} and are consequently more likely to disappear from the cohort before reaching more advanced age classes. The proportion of failed/damaged sub-systems thus tends to increase with age within an individual, but those individuals with more failures are systematically less likely to contribute to the observed distribution of failures in later age classes. In other words, the emergent distribution of failed sub-systems as a function of age in individuals is the result of a balance of forces on two levels --- failure accumulation within the organism and differential mortality risk between organisms --- acting in opposite directions. Thus, the progressive ageing of a cohort of organisms through the course of their lives can, in some sense, be viewed as multi-level selection in miniature; ageing individuals play the role of \posscite{luo_unifying_2014} collectives/groups/patches, and intra-organismal sub-systems involved in physiological function play the role of \posscite{luo_unifying_2014} particles/individuals.

 In this article, we make the above verbal argument precise. In section \ref{sec_bigpicture}, we provide a high-level verbal description of our general two-level stochastic model and the interpretation of its components. In section \ref{sec_definition}, we formally define the model as a continuous time Markov process with Moran-type dynamics simultaneously occurring at two levels. We then recast this into a stochastic ball-and-urn scheme following \citet{luo_unifying_2014} and carry out a system-size expansion/diffusion approximation~\citep{gardiner_stochastic_2009} to arrive at an approximate stochastic description. Section \ref{sec_scalinglimit} shows that the deterministic limit of our model recovers known models of actuarial senescence~\citep{gavrilov_biology_1991,woodbury_random-walk_1977,yashin_mortality_1985,bhat_failures_2026} and is equivalent to previous interpretations of the deterministic limit of \posscite{luo_unifying_2014} framework that appear when studying multi-level evolutionary games~\citep{cooney_replicator_2019,cooney_evolutionary_2023} and the evolution of cooperation in metapopulations~\citep{boorman_group_1973,kimura_evolution_1984,fontanari_groups_2014}. Our approach also results in dynamical, `pathwise' descriptions of the stochastic dynamics of the multi-level model in terms of stochastic (partial) differential equations and random fields/spacetime noise/Wiener processes, allowing for efficient simulation~\citep[Chapter 15]{gardiner_stochastic_2009} and the powerful tools of white noise and stochastic calculus~\citep{week_white_2021,bhat_stochastic_2025}.

\section{Methods: summary of the modelling approach}\label{sec_bigpicture}

We formulate a continuous time stochastic process that models selection at two levels inspired by \citet{luo_unifying_2014}. We call members of the lower level `\lows', and  members of the higher level `\highs', deliberately choosing rather abstract terms to highlight that the model is applicably to any two levels of biological organisation. In the case of senescence, `\low' corresponds to intra-organismal sub-systems (organs, organ systems, physiological processes, etc.) and `\high' corresponds to an individual organism; for metapopulations and group selection, the \low corresponds to an individual organism (within a patch/group) and the \high corresponds to a patch or group.

Our population contains $M$ \highs.  Within each \high, there are $N$ \lows that come in two types, \fails and \notfails. Again, we deliberately use the relatively abstract terms \fails and \notfails to avoid association with particular scenarios. In the case of senescence via damage/failure accumulation, failed sub-systems are \fails and functioning sub-systems are \notfails. In the case of evolution of cooperation, defectors are \fails and cooperators are \notfails. In the case of prudent metapopulations, \fails are non-prudent agents and \notfails are prudent agents. In all interpretations, the presence of \fails \lows increases the vulnerability of the \high entity that contains them, with the consequent increase of its rate of disappearance.

Within each \high, we will assume that the dynamics between \fails \lows and \notfails \lows is a birth-death process on $\{0,\frac{1}{N},\frac{2}{N},\cdots,1\}$. Additionally, each \high experiences a risk of disappearing from the collection, and we assume that this risk is dependent on the proportion of \fails \lows that the \high contains. When a \high disappears, with probability $\repprob$, it is instantaneously replaced by another \high chosen uniformly at random from from the set of surviving \highs. With probability $1-\repprob$, there is no such replacement. We will be particularly interested in the two extreme cases where replacement always occurs ($\repprob = 1$), in which case the model is analogous to a Moran model, and where replacement never occurs ($\repprob = 0$). In the case of senescence of individuals, disappearance corresponds to mortality. In the evolution of cooperation setting, disappearance refers to the extinction of a group and replacement can be thought of as `birth' at the group level. In the metapopulation setting, disappearance corresponds to (local) extinction of a patch, and replacement corresponds to re-colonisation of an empty patch by individuals of another patch in the style of infinite island models.

\section{The general stochastic description}\label{sec_definition}

Consider a collection of $M$ \highs, each comprised of $N$ \lows. We start observing this population at time $t=0$. For each $i \in \{1,2,\ldots,M\}$, let $F_i(t) \in \{0,1,2,\ldots,N\}$ denote the number of \fails \lows within the $i$\textsuperscript{th} \high at time $t > 0$. Our model is defined by specifying dynamics at both the `lower' level (between \lows that reside within the same \high) and the `higher' level (between different \highs).

\subsection{Lower-level dynamics}

Within a \high with $F \in \{0,1,2,\ldots,N\}$ \fails \lows, a \notfails \low turns into a \fails \low at rate $Nr_+(F/N)$, a process we call \failure. A \fails \low  turns into a \notfails \low at rate $Nr_-(F/N)$, a process we call \repair. Both \failure and \repair occur within each \high, at rates that depend only on the proportion of \fails \lows within the \high, and in particular, are independent of the state of every other \high in the collection. In other words, the dynamics within each \high is described by a continuous time Moran process with the transition rate parameters $Nr_{\pm}(F/N)$.

\subsection{Higher-level dynamics}

A \high with $F \in \{0,1,2,\ldots,N\}$ \fails \lows is chosen to disappear at rate $\mu(F/N)$. When a \high disappears, with probability $\repprob$, it is instantaneously replaced by a copy of another \high, chosen uniformly at random from the set of \highs that are currently present in the population. If $\repprob = 0$, no \high is ever replaced and we track a population of $M$ \highs until all of them have disappeared. If $\repprob = 1$, the total number of \highs always remains $M$ and the between \high dynamics are described by a Moran process with selection and Death-birth updating, or, equivalently, a biased voter model with Death-birth updating on the complete graph~\citep{yagoobi_update_2023}. We will call $\mu$ the `mortality hazard' and use `death' and `disappearance' interchangeably in anticipation of recovering models of senescence.
    

%

While the instantaneous replacement of all \highs ($\repprob = 1$) in higher-level dynamics may seem strange and un-biological for senescence (where \highs are individual organisms and \lows are intra-organismal sub-systems), we will see an interesting result where this replacement process produces probability distributions that describe the part of the cohort that is still alive (\emph{i.e.} the distribution conditional on remaining alive until a specified age). 

\subsection{Representation as a stochastic ball-and-urn process}

Having verbally described our stochastic process via both the lower and the higher level dynamics, we now seek a precise mathematical description. 
Here we follow \cite{luo_unifying_2014} by recasting our two-level process in terms of a ball-and-urn scheme (Fig \ref{fig_schematic}C). Consider $N$ `urns', one for each possible value of the proportion of \fails \lows within a \high, and $M$ `balls', one for each \high. The $i$\textsuperscript{th} ball (\high) is placed in the $(Nj)$\textsuperscript{th} urn if a proportion $j$ (out of $N$) of its \lows are \fails, and a ball (\high) is moved from one urn to another as the proportion of \fails \lows within the \high changes. If disappearance is with replacement, the total number of balls remains constant, whereas if there is no replacement, the total number of balls is initially $M$ and decreases over time. We will track the proportion of \highs out of $M$ that have $f = F/N$ \fails \lows. In the model with replacement, this is always equal to the proportion of \highs in the population that have $f$ \fails \lows (this is not so if there is no replacement because the total number of \highs changes over time). 

Concretely, we define the state of the stochastic process we are interested in at time $t$ as
\begin{linenomath*}\begin{equation}
\label{Moran_state_variable}
\xi^{M,N}(\cdot, t) \coloneqq \frac{1}{M}\sum\limits_{i=1}^{M}\delta_{f_{i}(t)}(\cdot) \quad,
\end{equation}\end{linenomath*}
where $f_i(t) \coloneqq F_i(t)/N$ and $\delta_x(\cdot)$ is a point mass at $x$, defined by
\begin{linenomath*}\begin{equation}
    \delta_x(y) \coloneqq \begin{cases}
        1 &; \ x=y\\
        0 &; \ \textrm{otherwise}
    \end{cases}
\end{equation}\end{linenomath*}
 Formally, $\xi^{M,N}$ is a mathematical object called a `finite measure'. Informally, for each $t \geq 0$, $\xi^{M,N}(\cdot,t)$ can be thought of as a function from $\{0,\frac{1}{N},\frac{2}{N},\ldots,1\}$, the set of possible \fails \lows within each \high, to the unit interval $[0,1]$. For each $f \in \{0,\frac{1}{N},\frac{2}{N},\ldots,1\}$, the quantity $\xi^{M,N}(f,t)$ returns the number of \highs in the collection that have $f$ \fails \lows at time $t$ divided by $M$, the initial number of \highs in the population. If replacement is guaranteed ($\repprob = 1$), the quantity $\xi^{M,N}(f,t)$ will equal the proportion of \highs in the population that have $f$ \fails \lows at time $t$, because the total number of \highs always remains $M$.

 \myfig{0.8}{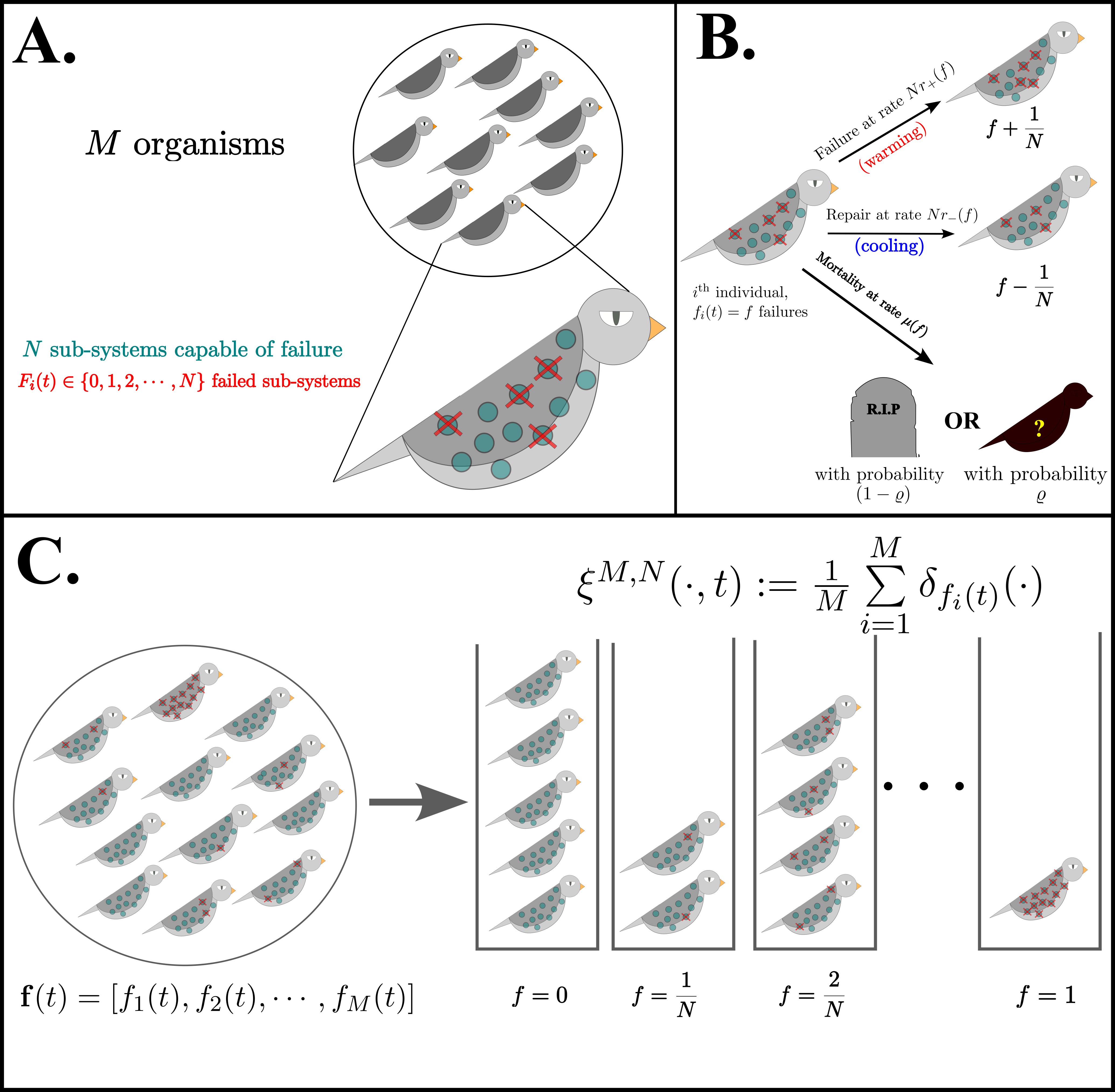}{A description of the stochastic process considered in this paper in the senescence setting, where \highs are organisms, \lows are intra-organismal sub-systems, \fails \lows are failed sub-systems, \failure is failure of functioning sub-systems, and \repair is repair of failed sub-systems. \textbf{(A)} The model consists of $M$ organisms, each of which have $N$ intra-organismal sub-systems capable of failure. \textbf{(B)} An organism with $f \in \{0,\frac{1}{N},\frac{2}{N},\ldots,1\}$ failures may either accumulate another failure, repair a failed sub-system, or die. Failure and repair follow lower-level dynamics. Death is either with instantaneous replacement (with probability $\repprob$) or without. \textbf{(C)} We characterise the population via a ball and urn scheme by counting the number of organisms that have $f = F/N$ failures for each possible value of $f$, and then dividing this number by $M$.}{fig_schematic}

One advantage of this abstract representation of the process is that for each $f$, the quantity  $\xi^{M,N}(f,t)$ always changes in units of $1/M$. In particular, we have now set up our process $\xi^{M,N}$ to jump via masses of unit $1/M$ that can take one of $N$ values. In other words, the size of the jumps is controlled by the total number of \highs ($M$), whereas the `granularity' of $\xi^{M,N}$ is controlled by the number of \lows within each \high ($N$). Henceforth, we use the notation $\Pi_N \coloneqq \{0,\frac{1}{N},\frac{2}{N},\frac{3}{N},\ldots,\frac{N-1}{N},1\}$ for $N < \infty$ to denote the $N$\textsuperscript{th} partition of the unit interval, the set of possible values of $f$. Notice that $\Pi_N \to [0,1]$ as $N \to \infty$, and so when the total number of \lows within a \high is infinite, $f$ varies continuously over $[0,1]$. In anticipation of taking this limit, we will use continuum notation throughout. Most importantly, this means we denote summation over states as $\int_{\Pi_N}df$ for both the discrete ($N < \infty$) and the continuous ($N \to \infty$) case.
In other words, $\int_{\Pi_N}df$ should be read as a stand-in for $\sum\limits_{f \in \Pi_N}$ when $N$ is finite. We stick to continuum notation to underscore the conceptual similarity between the discrete and continuous setting.

Our stochastic process $\{ \ \xi^{M,N}(\cdot,t) \ \}_{t\geq0}$ takes values in the space of sub-probability measures 
\begin{linenomath*}\begin{equation*}
 \mathcal{M}^{(M)}_{\leq 1}(\Pi_N) \coloneqq \left\{ \frac{1}{M}\sum\limits_{i=1}^{L}\delta_{x_i}  \quad \bigg{|} \quad x_i \in \Pi_N , L \leq M \right\}.   
\end{equation*}\end{linenomath*}
If replacement is certain ($\repprob = 1$), the state space is instead
\begin{linenomath*}\begin{equation*}
 \mathcal{M}^{(M)}_{1}(\Pi_N) \coloneqq \left\{ \frac{1}{M}\sum\limits_{i=1}^{M}\delta_{x_i}  \quad \bigg{|} \quad x_i \in \Pi_N  \right\},   
\end{equation*}\end{linenomath*}
the set of probability measures on $\Pi_N$. The elements of $\mathcal{M}^{(M)}_{1}(\Pi_N)$ can be thought of as probability mass functions when $N < \infty$ and as (discontinuous analogues of) probability density functions when $N = \infty$. When $\repprob < 1$, the total number of \highs decreases over time because some are lost to disappearance/mortality, and thus $\int_{\Pi_N}\xi^{M,N}(f,t)df < 1$ and $\xi^{M,N}$ does not describe a probability mass/density function (probability measure). 

Since the state space of our process is itself a set of distribution functions (measures), a distributional description answering the question  ``what is the probability of observing a population $\xi^{M,N}$ at time $t$?'' is asking about a distribution on a space of distributions. A more intuitive description would say how a given population $\xi^{M,N}$ is likely to change, since this can be visualised as changes to a probability distribution. We now provide such a `pathwise' description. 

\subsection{A pathwise stochastic description when the number of \highs is large}

 In supplementary section \ref{supp_sec_master_eqn}, we derive a Master Equation (Chapman-Kolmogorov equation) from the transitions verbally described above. We then carry out a system-size expansion (diffusion approximation) in the number of \highs ($M$) to derive an approximate expression for the change of $\xi^{M,N}$ in terms of expectations and variances. The diffusion approximation for $M$ is reasonable in situations where there are sufficiently many \highs that jumps of $\xi^{M,N}$ `look approximately continuous' at each fixed value of $f$.  

 We first introduce some simplifying notation. Let $D^{\pm}_N$ denote \emph{difference operators}, defined by their actions on functions $G: \Pi_N  \to \mathbb{R}$ as
\begin{linenomath*}\begin{equation}
\label{difference_operator}
    D^{\pm}_N[G(f)] \coloneqq N\left[G(f\pm\frac{1}{N})-G(f)\right]
\end{equation}\end{linenomath*}
%
In words, given a function $G$ evaluated at the point (proportion of \fails \lows) $f$, the difference operator $D^{\pm}_N[G(f)]$ tells us how much $G(f)$ differs from $G(f\pm\frac{1}{N})$, scaled by $N$. In the ball-and-urn language, given a function $G$ that is describing something about the $(Nf)$\textsuperscript{th} urn, $D^{\pm}_N[G(f)]$ tells us how different the $(Nf)$\textsuperscript{th} urn is from the  $(Nf \pm 1)$\textsuperscript{th} urn. Notice that these difference operators $D^{\pm}_N[G(f)]$ obey
\begin{linenomath*}\begin{equation}
\label{difference_operator_scaling}
    D^{\pm}_N[G(f)] \coloneqq N\left[G(f\pm\frac{1}{N})-G(f)\right] = \pm \frac{\partial G}{\partial f} + \frac{1}{2N}\frac{\partial^2G}{\partial f^2} + \mathcal{O}\left(N^{-2}\right),
\end{equation}\end{linenomath*}
and so converge to derivatives as $N$ grows larger, a fact that we will use later.
 
 In supplementary section \ref{sec_supp_diffusion_approx}, we show that when $M$ is not too small, our stochastic process satisfies the system of stochastic differential equations (a stochastic partial differential equation as $N \to \infty$) 
 \begin{linenomath*}\begin{equation}
\label{spde_maintext}
    \frac{\partial \xi^{M,N}}{\partial t} (f,t) = \mathcal{E}[f ,\xi^{M,N}] \ + \frac{1}{\sqrt{M}}\eta_{\xi^{M,N}}(f,t) \ ,
\end{equation}\end{linenomath*}
where
\begin{linenomath*}\begin{equation}
\begin{aligned}
\label{spde_mean}
\mathcal{E}[f,\xi^{M,N}] &\coloneqq  D^{-}_N\left[r_+(f)\xi^{M,N}(f,t)\right] + D^{+}_N\left[r_-(f)\xi^{M,N}(f,t)\right]\\
    &\hphantom{\coloneqq}- \left[  \ (1-\repprob)\mu(f) +  \repprob\left(\mu(f)-{\widehat{\mu}(t)}\right) \  \right] \  \xi^{M,N}(f,t). 
\end{aligned}
\end{equation}\end{linenomath*}
%
Here, we have introduced the scaled mean mortality in the population
\begin{linenomath*}\begin{equation}
\label{scaled_mean_mortality}
    \widehat{\mu}(t) \coloneqq \frac{\overline{\mu}(t)}{\ell(t)} \coloneqq \frac{\displaystyle \int\limits_{\Pi_N}  \mu(f)\xi^{M,N}(f,t)df}{\displaystyle \int\limits_{\Pi_N}  \xi^{M,N}(f,t)df}.
\end{equation}\end{linenomath*}
The numerator, $\overline{\mu}(t)$, weights each mortality $\mu(f)$ by the measure $\xi^{M,N}(f,t)$, the proportion of \highs out of $M$ that currently have $f$ \fails \lows. The denominator, $\ell(t)$, is the total proportion (out of $M$) of surviving \highs at time $t$, and thus the ratio $\widehat{\mu}$ is the mean mortality in the currently extant population. When $\ell(t) = 1$, the scaled mean mortality equals the mean mortality, $\widehat{\mu}(t) = \overline{\mu}(t)$. Notice that $\ell(t)$ equals 1 if and only if $\repprob = 1$ (i.e. every individual that disappears is guaranteed to be replaced).

The object $\eta_{\xi^{M,N}}(f,t)$ in Eq. \ref{spde_maintext} is variously called a \emph{spacetime noise process}, \emph{Gaussian random field}, \emph{stochastic field}, or $\mathcal{Q}$-\emph{Wiener process} on $\Pi_N\times [0,\infty)$ and captures the effects of stochastic fluctuations~\citep[Chapter 4]{daprato_spde_2014}. The noise is white (uncorrelated) in time and coloured (correlated) in (state) space. The covariance structure (`colour') of the noise process over states (\emph{i.e.} `between urns') is rather intricate, but serves largely as a book-keeping mechanism to track the correlations induced by replacement and the requirement that the total number of \lows within any \high always remains strictly constant. Since this covariance is ancillary to our results, we defer the precise definition of $\eta_{\xi^{M,N}}(f,t)$ to the supplementary (Eqs. \ref{FPE_diffusion}, \ref{spde_noise_mean}, and \ref{spde_noise_var}). For the main text, it is sufficient to note that the noise process always has zero expectation value, and thus $\mathcal{E}[f, \xi^{M,N}]$ alone describes the expected change in population composition over time. 

\section{Scaling limits}\label{sec_scalinglimit}

We now study the dynamics when the number of \highs in the collection ($M$) and the number of \lows within each \high ($N$) are large, and demonstrate that these scaling limits recover some previous descriptions of metapopulation dynamics~\citep{boorman_group_1973}, progression of senescence~\citep{gavrilov_biology_1991,yashin_mortality_1985}, and evolution of cooperation~\citep{kimura_evolution_1984,luo_unifying_2014}.

\subsection{Dynamics of infinitely many \highs ($M\to \infty, N\textit{ fixed)}$}\label{sec_deterministic_limit}

The noise process $\eta_{\xi^{M,N}}(f,t)$ in Eq. \ref{spde_maintext} is multiplied by $1/\sqrt{M}$. Thus, as the total number of \highs $M \to \infty$ for fixed $N$ (total number of \lows per \high), the stochastic fluctuations vanish and we obtain a deterministic process defined by an $N+1$ dimensional system of ordinary differential equations (ODEs), one for each $f \in \{0,\frac{1}{N},\frac{2}{N},\cdots,1\}$. 
Letting $P_{f}(t) \coloneqq \lim\limits_{M\to\infty}\xi^{M,N}(f,t)$ denote the proportion of \highs with $f$ \fails \lows in the $M \to \infty$ limit, Eq. \ref{spde_maintext} tells us that when $f \in \{\frac{1}{N},\frac{2}{N},\frac{3}{N},\cdots,\frac{N-1}{N}\}$, the limiting ODEs are
\begin{linenomath*}\begin{equation}
\label{moran_ODE_limit}
\frac{dP_f(t)}{dt} = D^-_N[r_+(f)P_f(t)] + D^+_N[r_-(f)P_f(t)] - \left((1-\repprob)\mu(f) + \repprob{[\mu(f) - \widehat{\mu}(t)]}\right)P_f(t)
\end{equation}\end{linenomath*}
At $f=0$ and $f=1$, the ODEs are instead
\begin{subequations}
\label{moran_ODE_limit_bdry_terms}
\begin{linenomath*}\begin{align}
    \frac{dP_0(t)}{dt} &= -Nr_+(0)P_0(t) + Nr_-\left(\frac{1}{N}\right)P_{\frac{1}{N}}(t) - \left[(1-\repprob)\mu(0) + \repprob(\mu(0                                                    ) - \repprob{\widehat{\mu}(t)}\right]P_0(t)\label{moran_ODE_limit_0_bdry}\\
    \frac{dP_1(t)}{dt} &= Nr_+\left(\frac{N-1}{N}\right)P_{\frac{N-1}{N}}(t) - Nr_-(1)P_1(t) - \left[(1-\repprob)\mu(1)+\repprob(\mu(1) - \repprob{\widehat{\mu}(t)})\right]P_1(t)\label{moran_ODE_limit_1_bdry}
    \end{align}\end{linenomath*}
\end{subequations}
Equations \ref{moran_ODE_limit} and \ref{moran_ODE_limit_bdry_terms} define a deterministic system of $N+1$ ODEs. If replacement is guaranteed ($\repprob = 1$), Eqs \ref{moran_ODE_limit} and \ref{moran_ODE_limit_bdry_terms} are a system of $N$ independent ODEs because $\sum_{k=0}^{N}\dot{P}_{k/N} = 0$. Eqs \ref{moran_ODE_limit}-\ref{moran_ODE_limit_bdry_terms} are similar to a `compartmental' structured population model (each value of $f$ being a compartment). 

\subsubsection{A classic senescence model with arbitrary finite $N$}

If we interpret \highs as individual organisms, \lows as intra-organismal sub-systems, \fails \lows as failed intra-organismal sub-systems, and \failure and \repair as failure and repair of sub-systems respectively, the dynamics predicted by these equations are identical to those studied in classic reliability theory models of senescence. In particular, if we set $r_-(f) = 0, r_+(f) = \lambda_0 + \lambda f, \mu(f) = \mu_0 + \mu f$ with positive constants $\lambda_0, \lambda, \mu_0, \mu$, the version of Eq. \ref{moran_ODE_limit} corresponding to the model without any replacement (i.e. $\repprob = 0$) recovers the model of `avalanche-like' accumulation of failures put forth by \cite{le_bras_lois_1976} and extended by \cite{gavrilov_biology_1991} (Section 6.4. Compare our Eq. \ref{moran_ODE_limit} with Eq. 71 in \cite{gavrilov_biology_1991}).

\subsection{Infinitely many \highs, each with a large number of \lows ($M \to \infty$ first, then $N$ large)}

As $N$ grows larger, the set $\{0,\frac{1}{N},\frac{2}{N},\cdots,1\}$ becomes the interval $[0,1]$ and $f$ starts to vary continuously. Thus, the system Eq. \ref{moran_ODE_limit} becomes an infinite system of ODEs --- in other words, a partial differential equation (PDE). Switching notation from $P_f(t)$ to $P(f,t)$ to switch from ODE to PDE notation, Eq. \ref{moran_ODE_limit} thus becomes the PDE
\begin{linenomath*}\begin{equation}
\label{moran_int_PDE_limit}
    \frac{\partial P(f,t)}{\partial t} =  D^-_N[r_+(f)P(f,t)] + D^+_N[r_-(f)P(f,t)] - \left[(1-\repprob)\mu(f) +\repprob\left(\mu(f) - \widehat{\mu}(t)\right)\right]P(f,t)
\end{equation}\end{linenomath*}
(with appropriate boundary conditions from Eq. \ref{moran_ODE_limit_bdry_terms}), and integrals are now over the unit interval,
\begin{linenomath*}\begin{equation}
\label{mean_mortality_continuum_limit}
    \widehat{\mu}(t) = \frac{\displaystyle \int\limits_{0}^{1}  \mu(x)P(x,t)dx}{\displaystyle \int\limits_{0}^{1} P(x,t)dx}.
\end{equation}\end{linenomath*}
%

\subsubsection{Infinitely many \highs, each containing a large but finite number of \lows}\label{sec_moran_KFPE_PDE_limit}

If we now assume $N$ is large but finite, we employ the \emph{diffusion approximation} from population genetics by setting $\frac{1}{N^2} \approx 0$. In this regime, defining the average \failure rate $r(f) \coloneqq r_{+}(f) - r_-(f)$ and the \low-level turnover $\tau(f) \coloneqq r_{+}(f) + r_-(f)$ and using Eq. \ref{difference_operator_scaling}, our Eq. \ref{moran_int_PDE_limit} becomes
\begin{linenomath*}\begin{equation}
\label{moran_KFPE_PDE_limit}
\resizebox{\textwidth}{!}{%
$\displaystyle  \frac{\partial P(f,t)}{\partial t} =  -\frac{\partial}{\partial f}\{r(f)P(f,t)\} + \frac{1}{2N}\frac{\partial^2}{\partial f^2}\{\tau(f)P(f,t)\} - \left[(1-\repprob)\mu(f) +\repprob\left(\mu(f) - \widehat{\mu}(t)\right)\right]P(f,t)
$
}
\end{equation}\end{linenomath*}
When \highs are patches/groups, \lows are individuals, and \fails \lows are non-cooperators, Eq. \ref{moran_KFPE_PDE_limit} with $\repprob = 0$ is Eq. 3 in \cite{boorman_group_1973}, and with $\repprob = \ell(t) = 1$ is Eq. 2.5 in \cite{kimura_evolution_1984} and Eq. 11 in \cite{fontanari_groups_2014}. In these models, \failure corresponds to a defector replacing a cooperator, \repair corresponds to a cooperator replacing a defector, and disappearance corresponds to extinction (followed by recolonisation when there is replacement).

When `\lows' are intra-organismal sub-systems, `\highs' are individual organisms and \failure is failure of a sub-system,  Eq. \ref{moran_KFPE_PDE_limit} without replacement ($\repprob = 0$) is an extension of the Woodbury-Manton model of senescence~\citep{woodbury_random-walk_1977,bhat_failures_2026} whereas Eq. \ref{moran_KFPE_PDE_limit} with guaranteed replacement ($\repprob = \ell(t) = 1$) describes the dynamics of the same model when we track the conditional probability $P_c(f,t)$ of an individual of age $t$ having $f$ failures, conditioned on the individual remaining alive up until this age~\citep{yashin_mortality_1985,bhat_failures_2026}. 

A minimal, biologically motivated choice of functional form for $r$ and $\mu$ in this setting produces Gompertz-Makeham mortality curves together with the possibility of late-life mortality deceleration due to selective disappearance~\citep{bhat_failures_2026}.  Remarkably, though Eq. \ref{moran_KFPE_PDE_limit} describes an infinitely large cohort in which each individual is always instantaneously replaced upon death, the result recapitulates earlier studies that formulate stochastic models~\citep{yashin_mortality_1985,bhat_failures_2026} tracking the probability distribution of failures within a \emph{single} focal individual, conditioned on that individual \emph{not yet having died}. 

This correspondence can intuitively be understood by observing that instantaneous replacement by an individual chosen uniformly at random from the currently living population amounts to a form of `unbiased renormalisation': When we condition on individuals remaining alive, we are effectively saying ``only focus on those individuals who remain alive''. Recalling the every day definition of probability, we would want to calculate 
\begin{equation*}
    \mathbb{P}(\text{age $t$ individual has $f$ failures}) = \frac{\text{Number of (living) age $t$ individuals with $f$ failures}}{\text{Number of individuals alive at age $t$}}
\end{equation*}
However, $\xi^{M.N}$ and thus $P(f,t)$ in Eq. \ref{moran_KFPE_PDE_limit} calculates the proportion of individuals of age $t$ out of the \emph{initial cohort size}. If we start with a cohort of a fixed number of individuals of age 0, the number of individuals who are alive at age $t$ decreases as $t$ increases. Thus, the denominator of the above expression changes over time, or, in other words, the number of ``probability slots'' we are calculating our probabilities over decreases with time/age. We thus need to dynamically ``renormalise'' the distribution of failures at each time (keep track of the denominator of the expression) to make sure the probability distribution of failures over the set of extant individuals still sums/integrates to 1. Replacement when $\repprob = 1$ automatically does this by letting each still alive individual replace the ``probability slot'' of an individual who dies. This does not change the distribution of failures because every living individual is assumed to be equally likely to contribute to the ``probability slot'', but ensures that the number of ``probability slots'' remains unchanging over time. 

\subsubsection{Infinitely many \highs, each containing infinitely many \lows}\label{sec_moran_replicator_PDE_limit}

If we instead take $N \to \infty$ in Eq. \ref{moran_int_PDE_limit}, we once again use Eq. \ref{difference_operator_scaling} to see that the resulting PDE is
\begin{linenomath*}\begin{equation}
\label{moran_replicator_PDE_limit}{
    \frac{\partial P(f,t)}{\partial t} =  -\frac{\partial}{\partial f}\{r(f)P(f,t)\} - \left[(1-\repprob)\mu(f) +\repprob\left(\mu(f) - \widehat{\mu}(t)\right)\right]P(f,t)
}
\end{equation}\end{linenomath*}
When replacement is guaranteed ($\repprob = \ell(t) = 1$, and thus $\widehat{\mu} = \overline{\mu} = \int_0^1\mu(f)P(f,t)df$), our Eq. \ref{moran_replicator_PDE_limit} is a version of \posscite{luo_unifying_2014} deterministic limit for her choice of functions $r_\pm$ and $-\mu$. Equation \ref{moran_replicator_PDE_limit} in the multi-level selection context has been interpreted as a two-level replicator equation and extended to a variety of evolutionary games~\citep{cooney_replicator_2019,cooney_evolutionary_2023}. To make the connection between our equation and these results exact, observe that mortality can be interpreted as a `negative payoff' and compare our Eq. \ref{moran_replicator_PDE_limit} with Eq. 5 in \citet{cooney_evolutionary_2023} with $\varrho = \ell = 1$ and the choice $r(f) = f(1-f)(\pi_C(f)-\pi_D(f))$ (their equation in turn recovers Eq. 1 in \cite{luo_unifying_2014} as a special case). The first term on the RHS of Eq. \ref{moran_replicator_PDE_limit} represents the effects of selection within a \high (\emph{i.e.} between \lows), whereas the second term on the RHS represents the effects of selection between \highs. In the senescence models, these respectively correspond to the progression of failures of sub-systems within an individual, and  differential mortality risk causing selective disappearance, a between-individual process~\citep{yashin_mortality_1985,bhat_failures_2026}.  

\subsection{A stochastic limit when there is high extrinsic mortality (high group-level extinction but weak group-level selection) and $N$ and $M$ are comparable}

Throughout this article, our population has experienced \low level events (\failure and \repair) at $\mathcal{O}(MN)$ rates and \high level events (disappearance and replacement) at $\mathcal{O}(M)$ rates. Such scaling naturally leads to within-\high dynamics unfolding more rapidly than between-\high dynamics. This is a sensible setup when the between-\high events (disappearance/replacement) depend only on the proportion of \fails \lows they contain, intuitively because each of the $M$ \highs contains $N$ \lows. As an aside, at this point we deviate from our usual praise of \posscite{luo_unifying_2014} work: while technically correct, their analysis included a parameter $w$ that determined the relative timescales of between-\high and within-\high events; while it is mathematically fine to analyse the consequences of all values of $w$, their decision to consider the situation where between-\high and within-\high events are on exactly the same timescale ($w=1$ in their notation) as a kind of `default' creates an inadvertent emphasis on fast group-level phenomena in a way that many researchers participating in the group selection debate would probably consider biologically unlikely.

In the context of senescence, however, biologically plausible rates may be different. One expects to occasionally encounter situations where individuals  experience a high level of mortality that is due to chance events and independent of their state (proportion of failed sub-systems), called `extrinsic mortality' in senescence lingo~\citep{moorad_2019_extrinsic,de_vries_extrinsic_2023}. In some situations such as for organisms in harsh or predator-rich environments, extrinsic mortality can  occur at a rate that renders the timescales of disappearance and replacement to be comparable to those of \failure and \repair, in which case our original scaling assumption would fail.


In this section, we study an alternative scaling regime by assuming that the disappearance rate is of the form $\mu(f) = N\mu_e + \nu(f)$, where $\mu_e > 0$ is a constant baseline disappearance rate and $\nu(f)$ is an $\mathcal{O}(1)$ term representing state-dependent disappearance as before. In multi-level selection language, this is a weak selection assumption --- most between-\high dynamics are due to the composition-independent baseline $N\mu_e$, with the relatively small additional component $\nu(f)$ representing the additional vulnerability to disappearance caused by the presence of \fails \lows, causing between-\high selection. 


If $M$ and $N$ are both small, the dynamics defined by the process are analytically intractable. Since we are interested in what happens when $M$ and $N$ are comparable to each other rather than small in absolute terms, we take the limit $N\to \infty, M \to \infty$ together such that their ratio $N/M \coloneqq \lambda$ remains fixed. Since the mortality of \highs is now also scaled by $N$,  all $M$ \highs that we start with will very quickly disappear if $\repprob$ is small in this regime. To see non-trivial dynamics, we will thus further assume $\repprob = 1 - (\varepsilon/M)$, where $\varepsilon$ quantifies the (small) probability that a \high that disappears is not replaced. 


Let $\zeta^\lambda(f,t)$
%
%
denote the limiting process. In supplementary section \ref{supp_sec_superprocess}, we show that $\zeta^{\lambda}$ is described by the stochastic partial differential equation (SPDE)
\begin{linenomath*}\begin{equation}
\label{superprocess_spde}
    \frac{\partial \zeta^{\lambda}}{\partial t}(f,t) = \mathcal{E}_{\lambda}[f,\zeta^{\lambda}] +  \sqrt{\mu_e\lambda} \ \eta_{\zeta^{\lambda}}(f,t),
\end{equation}\end{linenomath*}
where
\begin{linenomath*}\begin{equation}   
    \mathcal{E}_{\lambda}[f,\zeta^{\lambda}] = -\frac{\partial}{\partial f}\left\{r(f)\zeta^\lambda(f,t)\right\} - \bigg[\varepsilon\lambda\mu_e + \bigg(\nu(f)-\widehat{\nu}(t)\bigg)\bigg]\zeta^{\lambda}(f,t) \label{superprocess_spde_mean}
\end{equation}\end{linenomath*}
and $\eta_{\zeta^{\lambda}}(f,t)$ is a spacetime noise process (Gaussian random field) that is white in time but coloured in state space, with a covariance structure presented in supplementary section \ref{supp_sec_superprocess}. While both the `lower'-level processes (\failure and \repair) and the `higher'-level processes (disappearance and replacement) affect the expected behaviour of the SPDE via Eq. \ref{superprocess_spde_mean}, supplementary section \ref{sec_superprocess_var} proves that lower-level dynamics do not affect the (co)-variance and thus the stochastic fluctuations of the process. This is intuitive, because \failure and \repair alter the state of a single \low, but disappearance and replacement alter the fate of all $N$ \lows contained within the focal \high in one swoop, and thus the latter events cause more pronounced stochasticity in population composition. The strength of the noise in Eq. \ref{superprocess_spde} is controlled by $\mu_e$, the baseline mortality rate common to all \highs, and $\lambda = N/M$, a measure of the relative ``population sizes'' at the two levels. If the number of \lows $M$ is much larger than the number of \lows per \high $N$, the ratio $\lambda \to 0$ and Eq. \ref{superprocess_spde} reduces to the deterministic Eq. \ref{moran_replicator_PDE_limit} with guaranteed replacement ($\repprob = \ell(t) = 1$) that has been the focus of multi-level selection literature (e.g. \citet{cooney_replicator_2019,cooney_evolutionary_2023}).

Eq. \ref{superprocess_spde} describes a type of stochastic process called a ``superprocess''~\citep{etheridge_2000_superprocesses} that takes values in the set of sub-probability measures on $[0,1]$. When replacement is certain ($\varepsilon = 0, \ell(t) = 1$), the covariance structure of the superprocess is exactly that of a population genetic infinite-allele model called the Fleming-Viot process~\citep{fleming_process_1979,etheridge_2000_superprocesses} that generalises the more classical Wright-Fisher and Moran processes~\citep{etheridge_2000_superprocesses}. \citet{luo_scaling_2017} have also rigorously proven convergence to a Fleming-Viot type process in a similar, albeit not identical, scaling limit (\citet{luo_scaling_2017} assume `within-\high' selection is also weak by setting $r_+ - r_-$ to be $\mathcal{O}(1/N)$, and as a consequence, need to additionally rescale time; they also exclusively work with $\varepsilon = 0$). When $\varepsilon > 0$, our superprocess is no longer strictly Fleming-Viot because $\zeta^{\lambda}$ no longer describes a probability distribution.

\section{Discussion}

We have shown that the ageing of a cohort of individuals over the course of their lifetimes is formally equivalent to a multi-level selection process, with progressive deterioration of physiological function due to interdependencies (``failure begets failure'', \citet{bhat_failures_2026}) playing the role of within-group selection, and selective disappearance due to differential mortality risk playing the role of between-group selection. We used a framework very similar to \posscite{luo_unifying_2014} model of multi-level selection, with some generalisations and changes, as discussed below. We have also shown that as the total number of collectives tends to infinity, conditioning on having survived (\emph{i.e.} asking questions about the part of a cohort still alive) in the senescence setting becomes formally equivalent to asserting that local extinction is immediately accompanied by recolonisation in the metapopulation setting, or that the within-group process follows a constant population Moran process in the multi-level evolutionary game and group selection setting.

One major upshot of making the formal connection between senescence (within a generation) and multi-level selection is that it allows us to translate statements about one class of models to statements about the other. For instance, \citet{cooney_evolutionary_2023} have recently synthesised the ways in which within-group and between-group processes can promote the persistence of cooperation in the deterministic limit of (extensions of) \posscite{luo_unifying_2014} framework. Their equation is a special case of ours in this limit (set $\repprob = \ell(t) = 1, r(f) = f(1-f)(\pi_C(f)-\pi_D(f))$ in our Eq. \ref{moran_replicator_PDE_limit} and compare with Eq. 5 in \cite{cooney_evolutionary_2023}, remembering that mortality is a negative payoff). Thus, translating from \posscite{cooney_evolutionary_2023} multi-level evolutionary games to \posscite{yashin_mortality_1985} (and \posscite{bhat_failures_2026}) demographic senescence, the routes to cooperation described by \citet{cooney_evolutionary_2023} could provide insights into mechanisms through which different intra-organismal interdependency structures~\citep{gavrilov_biology_1991} and mortality dynamics may reduce the overall rate of physiological deterioration and prolong healthspan, an important aspect of senescence~\citep{williams_1999_tithonus}. 

Why hasn't earlier work uncovered the equivalence we describe here? We believe the answer likely rests in the stark terminological differences between the fields of multi-level selection and senescence research. In the former, one typically speaks of cooperators `creating a collective advantage' at the between-group level, or defectors `exploiting' cooperators at the within-group level~\citep{okasha_MLSbook_2006,okasha_2018_agents}, potentially leading to a `tragedy of the commons'~\citep{rankin_2007_tragedy}. This active, agential language is at odds with phrases describing senescence in reliability theory: ageing is spoken of as a consequence of `accumulation of damage' or the `risk of mortality due to deterioration'~\citep{gavrilov_biology_1991,ledberg_exponential_2020}, distinctly passive phrases. Our work shows that this verbal distinction, while useful for back-of-the-envelope reasoning, has no deeper meaning: the formal descriptions of senescence within a generation (e.g. \cite{woodbury_mathematical_1983,yashin_mortality_1985}) and group selection (e.g. \cite{luo_unifying_2014,cooney_evolutionary_2023}) both obey Eq. \ref{moran_KFPE_PDE_limit} and hence are equivalent. 

One who enjoys agential language could say that physiological sub-systems, such as organs, that cooperate to keep an organism alive enjoy a cohort-level advantage of existing at later ages (for a discussion on the validity of agential language in these fields, see \citet{okasha_2018_agents,agren_agency_2022}). Equivalently, cooperation research could employ passive language with defectors accumulating within a focal group, which increases the vulnerability of the group to going extinct in multi-level selection models. It is tempting to suggest that the existing differences in agential language reflect the fact that humans feel capable of consciously choosing whether to cooperate in a given setting (whether or not we have free will or agency, the \textit{feeling} is that we do, \cite{dennett_2001_hunch,fischer_four_2024}), while few would claim to have the same feeling about being able to control the decay of one's own ageing body by willpower alone.

Our model also extends and complements \posscite{luo_unifying_2014} model and its successors. First, we do not assume a particular functional form for either the \low-level dynamics ($r_\pm$; `individual-level fitness' in \citet{luo_unifying_2014}) or the \high-level dynamics ($-\mu$; `group-level fitness' in \citet{luo_unifying_2014}). Second, we have incorporated a parameter $\repprob$ that allows us to relax the assumption that the total number of \highs is strictly constant. Instead, in our model, the number of extant \highs is non-increasing over time, and decreasing over time whenever $\repprob < 1$ and $\mu \not\equiv 0$. In the context of metapopulation dynamics, $\repprob < 1$ means that not every patch that goes locally extinct is recolonised, which could be more realistic in systems such as island chains where recolonisation and re-establishment following local extinction is not guaranteed. Third, while previous extensions of \posscite{luo_unifying_2014} framework~\citep[e.g. ][]{luo_unifying_2014,luo_scaling_2017,vanveelen_groups_2014,cooney_replicator_2019,cooney_2022_chromosomes,cooney_long-time_2022,cooney_evolutionary_2023} have incorporated heterogeneity between entities in the birth rate, we have ascribed them to death rates instead --- a more natural choice for senescence. The difference is relevant because fitness effects in many scenarios affect survival rather than fecundity, and in metapopulation contexts in particular, regulate vulnerability to local extinction. Since model outcomes can be sensitive to whether fitness alters the birth rate or the death rate in the stochastic setting~(see e.g. \cite{mcleod_social_2019,raatz_promoting_2023}), our model could differ from these previous studies in the stochastic regime, though our framework is equivalent to theirs in all deterministic limits.

When extrinsic mortality is high (group level events are frequent but group selection is weak), we have shown that the dynamics of our model are not deterministic, but instead described by a stochastic process called a `superprocess' that solves a stochastic partial differential equation (Eq. \ref{superprocess_spde}). The behaviour of the model in this regime may be mathematically interesting in its own right as a biologically motivated superprocess that has non-trivial deterministic components and resembles a Fleming-Viot process for some ($\varepsilon = 0$) but not all ($\varepsilon > 0$) natural parameter values. A major research theme around Fleming-Viot superprocesses investigates duality with a backwards-in-time process called the coalescent, with applications in population genetic inference~\citep{birkner_FV_2009}. Since Eq. \ref{superprocess_spde} is almost a Fleming-Viot process, a compelling  mathematical challenge is to address whether duality with a coalescent can be established for our process (which has a non-trivial, frequency-dependent directional/``selection'' component due to Eq. \ref{superprocess_spde_mean}), and whether such a duality can be extended to the $\varepsilon > 0$ case where population size decreases over (forward) time. In the context of senescence, duality with a coalescent would potentially allow one to make inferences about the past history of physiological conditions (damage/failures) that an individual of known age and condition is likely to have experienced.

Our work provides pathwise descriptions of stochastic multi-level selection in terms of stochastic (partial) differential equations (Eq. \ref{spde_maintext}, Eq. \ref{superprocess_spde}). It thus enables studying the effects of having a small number of groups in multi-level selection models such as \citet{vanveelen_groups_2014} and \citet{cooney_evolutionary_2023}, and the effects of limited cohort sample size in senescence models such as \citet{yashin_mortality_1985} and \citet{bhat_failures_2026}. The equations we derive are also amenable to attack using a diverse analytical toolbox comprised of stochastic calculus~\citep{week_white_2021,bhat_stochastic_2025}, phase transition theory~\citep[Chapters 3-5]{garcia-ojalvo_noise_1999}, and timescale separation via slow-manifold approximation~\citep{constable_2013_manifold,constable_2014_fast,parsons_2017_manifold}. For instance, the stochastic dynamics of Eq. \ref{spde_maintext}, as well as of the more challenging superprocess Eq. \ref{superprocess_spde}, are formulated both as stochastic (partial) differential equations (Eq. 
\ref{spde_maintext} and Eq. \ref{superprocess_spde}) and equivalently as (functional) Fokker-Planck equations (Eq. \ref{functional_FPE} and Eq. \ref{superprocess_FPE}). This means the stochastic dynamics, complete with the coloured noise, can be analytically studied using both deterministic PDE tools such as spectral methods (see e.g. chapters 8, 9, 12, and 13 of \citet{gardiner_stochastic_2009}) and SDE/SPDE tools such as quasi-stationary densities/speed measures~\citep{czuppon_understanding_2021} and It\^o stochastic calculus (see supplementary section 4.2 in \cite{week_gmatrix_2026} for powerful and immediately applicable heuristics in the stochastic calculus vein. Week applies these heuristics to study the effects of genetic correlations as captured by $G$-matrices in affecting multi-variate evolution in finite populations). The equations we derive are hence promising starting points for the analytical study of the stochastic aspects of both multi-level selection and senescence via failure accumulation.

Lastly, the senescence process we model here only describes how senescence progresses through chronological age in a cohort of individuals (or, equivalently, within the typical individual as $M \to \infty$). This means that our result, that senescence progression is (in a formal sense) a version of multi-level \emph{selection} (in the sense of \citet{luo_unifying_2014,cooney_evolutionary_2023}) , is thus far restricted to processes operating within a generation. How senescence \emph{evolution} might be related to multi-level \emph{evolution} is beyond our current scope. We consider this an exciting avenue for future research.


\section*{Acknowledgements}\addcontentsline{toc}{section}{Acknowledgements}

The GenEvo RTG funded by the Deutsche Forschungsgemeinschaft (DFG, German Research Foundation) – GRK2526/1 – Project nr. 407023052 is gratefully acknowledged.
	
\section*{Author Contributions}
\textbf{Ananda Shikhara Bhat:} Conceptualisation, Methodology, Formal Analysis, Writing - Original Draft, Writing - Review \& Editing, Visualisation; \textbf{Hanna Kokko:} Validation, Writing - Review \& Editing, Supervision.

\begin{refcontext}[sorting=nyc]
\printbibliography[title=References]\addcontentsline{toc}{section}{References}
\end{refcontext}
\end{refsection}

\begin{refsection}
\nocite{ethier_1986_markov,pawula_approximation_1967,rudin_functional_1991}
\begin{refcontext}[sorting=nyc]
\printbibliography[title=References Cited Only in the Supplementary]
\end{refcontext}
\end{refsection}


\clearpage
\newpage
\nolinenumbers
\begin{center}
	{\LARGE \bfseries Supplementary Information for \\
		Bhat and Kokko: 
        Demographic senescence as multi-level selection in miniature\\
		\vspace{2em}
		\normalfont\Large Ananda Shikhara Bhat\textsuperscript{1,2,$\ast$} and Hanna Kokko\textsuperscript{1,2,$\dag$}\\
		\vspace{3em}

\noindent{} 1. Institute of Organismic and Molecular Evolution (iomE), Johannes Gutenberg University, 55128 Mainz, Germany;

\noindent{} 2. Institute for Quantitative and Computational Biosciences (IQCB), Johannes Gutenberg University, 55128 Mainz, Germany;
\vspace{3em}\\
\noindent{} $\ast$ E-mail: abhat@uni-mainz.de\\
\noindent{} $\dag$ E-mail: hkokko@uni-mainz.de
}
\end{center}
\newpage


\renewcommand{\theequation}{S\arabic{equation}}
\renewcommand{\thetable}{S\arabic{table}}
\renewcommand{\thesection}{S\arabic{section}}
\renewcommand{\thefigure}{S\arabic{figure}}
\setcounter{figure}{0}
\setcounter{table}{0}
\setcounter{section}{0}
 \setcounter{equation}{0}  
\pagestyle{fancy}
\rhead{Supplement to Bhat and Kokko, 
       \textit{Demographic senescence as multi-level selection}
       }
\setlength{\headsep}{0.3in}
\setlength{\headheight}{14pt}  
\lhead{} 


\resetlinenumber
\linenumbers

\begin{refsection}


\section{Deriving our main SPDE (Eq.  \ref{spde_maintext})}\label{supp_sec_spde_derivation}

From the lower-level dynamics and higher-level dynamics, our ball-and-urn representation $\xi^{M,N}$ can change in the following ways for each $f \in \Pi_N$:
 \begin{enumerate}
     \item \textbf{\failure} moves a ball one urn to the right (from $f$ to $f + \frac{1}{N}$). This changes $\xi^{M,N}$ to $\xi^{M,N} - \frac{1}{M}\delta_f + \frac{1}{M}\delta_{f+\frac{1}{N}}$. There are $M\xi^{M,N}(f)$ individuals that have $f$ \fails \lows, and each experiences \failure at a rate $Nr_+(f)$. Thus, this event occurs at total rate $Nr_+(f)M\xi^{M,N}(f)$.
     \item \textbf{\repair} moves a ball one urn to the left (from $f$ to $f - \frac{1}{N}$). This changes $\xi^{M,N}$ to $\xi^{M,N} - \frac{1}{M}\delta_f + \frac{1}{M}\delta_{f-\frac{1}{N}}$. There are $M\xi^{M,N}(f)$ individuals that have $f$ \fails \lows, and each experiences \repair at a rate $Nr_-(f)$. Thus, this event occurs at total rate $Nr_-(f)M\xi^{M,N}(f)$.
     \item \textbf{Disappearance without replacement} removes a ball from an urn. This changes $\xi^{M,N}$ to $\xi^{M,N} - \frac{1}{M}\delta_f$. There are $M\xi^{M,N}(f)$ individuals that have $f$ \fails \lows, and each experiences a mortality hazard of $\mu(f)$. There is no replacement with probability $(1 - \repprob )$. Thus, this event occurs at a total rate $(1-\repprob )M\mu(f)\xi^{M,N}(f)$.
     \item \textbf{Disappearance with replacement} removes a ball from an urn, and places a ball in another (or possibly the same) urn. A replacement event in which a \high with $k$ \fails \lows is replaced by a \high with $f$ \fails \lows changes $\xi^{M,N}$ to $\xi^{M,N} - \frac{1}{M}\delta_k + \frac{1}{M}\delta_{f}$. Death of a \high with $k$ \fails \lows occurs at total rate $M\mu(k)\xi^{M,N}(k)$; since the replacing \high is chosen uniformly at random from the population of extant \highs, replacement by a \high with $f$ \fails \lows occurs with probability $\xi^{M,N}(f)/(\int_{\Pi_N}\xi^{M,N}(f)df)$. There is replacement with probability $\repprob$. Thus, if we introduce the survivorship $\ell \coloneqq \int_{\Pi_N}\xi^{M,N}(f)df$, the replacement of a \high with $k$ \fails \lows by one with $f$ \fails \lows happens at total rate $ \ell^{-1}\repprob M\mu(k)\xi^{M,N}(k)\xi^{M,N}(f)$.
 \end{enumerate}
Thus, for each $f, k \in \{0,\frac{1}{N},\frac{2}{N},\ldots,1\}$, our stochastic process has the infinitesimal transition rates
\begin{subequations}
\label{transition_rates}
    \begin{linenomath*}\begin{align}
        \xi^{M,N} &\to \xi^{M,N} - \frac{1}{M}\delta_f +\frac{1}{M}\delta_{f+\frac{1}{N}} \ &&\textrm{ at rate } \ \mathcal{L}_+(f | \xi^{M,N}) \quad \text{(\failure)} \label{failure_rate}\\
        \xi^{M,N} &\to \xi^{M,N} - \frac{1}{M}\delta_f + \frac{1}{M}\delta_{f-\frac{1}{N}} \ &&\textrm{ at rate } \ \mathcal{L}_-(f | \xi^{M,N}) \quad \text{(\repair)}\\
        \xi^{M,N} &\to \xi^{M,N} - \frac{1}{M}\delta_f  \ &&\textrm{ at rate } \ \mathcal{D}(f | \xi^{M,N}) \quad \text{(disappearance, no replacement)} \\
        \xi^{M,N} &{\to \xi^{M,N} + \frac{1}{M}\delta_f} - \frac{1}{M}\delta_k \ && {\textrm{ at rate } \ \mathcal{R}(k, f | \xi^{M,N}) \quad \text{(replacement)}}
    \end{align}\end{linenomath*}
\end{subequations}
where
\begin{subequations}
\label{transition_functional_forms}
    \begin{linenomath*}\begin{align}
        \mathcal{L}_+(f|\xi^{M,N}) &\coloneqq MNr_+(f)\xi^{M,N}(f)\\
        \mathcal{L}_-(f|\xi^{M,N}) &\coloneqq MNr_-(f)\xi^{M,N}(f)\\
        \mathcal{D}(f|\xi^{M,N}) &\coloneqq M(1-\repprob )\mu(f)\xi^{M,N}(f)\\
        {\mathcal{R}(k,f|\xi^{M,N})} &{\coloneqq M\ell^{-1}\repprob \mu(k)\xi^{M,N}(k)\xi^{M,N}(f)}
    \end{align}\end{linenomath*}
\end{subequations}

The notation $(f  | \xi)$ is meant to emphasize that the quantities describe the transition rates \emph{at the point (\high with a proportion of \fails \lows)} $f$ \emph{in a state (population of \highs) } $\xi$. The replacement rate $\mathcal{R}$ involves two points, $k$ and $f$, corresponding to the \high that disappeared and the \high that replaced it respectively. We now use these infinitesimal rates to characterise our stochastic process in terms of a functional Master equation and its diffusion approximations. 

\subsection{Deriving a master equation}\label{supp_sec_master_eqn}

Let $P(\xi^{M,N},t)$ denote the probability of observing a collection described by $\xi^{M,N}$ at time $t$ (we assume this exists even when $N = \infty$, \emph{i.e.} that the process always admits a density). The probability $P(\xi^{M,N},t)$ is, of course, really conditioned on an initial state $\xi_0$, but we suppress this dependence on initial condition for conciseness. Notice that the second argument in the operators is an element of the state space of our stochastic process. In every case, $\xi$ can be viewed as an element of a metric space. In particular, $\xi^{M,N}$ can be thought of as an $N$-dimensional vector of proportions (one for each possible value of $f$) when $N$ is finite, and as a ``function'' (sub-probability measure) on $[0,1]$ when $N$ is infinite. We do not notationally distinguish between finite and infinite $N$ throughout this section because of the clear conceptual equivalence, and use continuum notation (integrals, PDEs, etc) throughout. These objects should be interpreted as stand-ins for the natural discrete analogues (sums, ODEs, etc) when $N$ is finite. We now follow the general notation and strategy of \citet{bhat_stochastic_2025}, describing $P(\xi^{M,N},t)$ in terms of a (functional) Master equation and then using (an infinite-dimensional analogue of) the diffusion approximation to arrive at a (functional) Fokker-Planck equation, or, equivalently, a stochastic (partial) differential equation.

To derive our master equation, we first require some notation. For each $k \in \Pi_N$, let $\mathcal{S}^{\pm}_{k,M}$ denote a step operator that, given any (nice) functional $G: \Pi_N \times \mathcal{M}^{(M)}_{\leq1} (\Pi_N) \to \mathbb{R}$, shifts it by $\pm \delta_h/M$: 
\begin{linenomath*}\begin{align*}
    \mathcal{S}^{\pm}_{k,M}G(f,\xi) &\coloneqq G\left(f , \xi \pm \frac{1}{M}\delta_k \right).
\end{align*}\end{linenomath*}
In words, given a functional $G(f,\xi)$, the shifted functional $\mathcal{S}^{\pm}_{k,M}G$ tells us what the quantity would have been at the same point $f$ if the population was shifted by $\xi \pm \frac{1}{M}\delta_k$ (i.e. if there was either one more ($+$) or one fewer ($-$) ball (\low) in the $(Nk)$\textsuperscript{th} urn).

We can now use the rates Eq. \ref{transition_rates} to count all possible transitions into and out of a focal state $\xi^{M,N}$, with the goal of writing down a master equation for $P(\xi^{M,N},t)$ in terms of the probability flux through the state $\xi^{M,N}$:
\begin{linenomath*}\begin{equation*}
    \frac{\partial P(\xi^{M,N},t)}{\partial t} = \text{(Rate of inflow of probability into $\xi^{M,N}$)} - \text{(Rate of outflow of probability from $\xi^{M,N}$)}.
\end{equation*}\end{linenomath*}
As an illustration of how one writes down the master equation, we compute the contributions of \failure to the probability flux through the state $\xi^{M,N}$. From the transition Eq. \ref{failure_rate}, we see that \failure transforms the state $\xi^{M,N} + \frac{1}{M} \delta_f - \frac{1}{M}\delta_{f + \frac{1}{N}}$ into the state $\xi^{M,N}$. Since this can happen for every $f \in \Pi_N$, the total `inflow' of probability into $\xi^{M,N}$ due to \failure is given by
\begin{linenomath*}\begin{equation}
\begin{aligned}
\label{rate_in_lower_below}
R^{\textrm{\failure}}_{\textrm{in}}(\xi^{M,N}, t) &= \underbrace{ \int\limits_{\Pi_N} }_{\substack{\text{`sum over'}\\\text{all possible $f$}}} \underbrace{\mathcal{L}_+(f | \xi^{M,N} + \frac{1}{M} \delta_f - \frac{1}{M}\delta_{f + \frac{1}{N}}) }_{\substack{\text{Rate of}\\ {(\xi^{M,N} + \frac{1}{M} \delta_f - \frac{1}{M}\delta_{f + \frac{1}{N}}) \to \xi^{M,N}} \\ \text{transition due to \failure} }}\ \underbrace{P(\xi^{M,N} + \frac{1}{M} \delta_f - \frac{1}{M}\delta_{f + \frac{1}{N}}, t)}_{\substack{\text{Probability of}\\\text{finding the state} \\ \text{$\xi^{M,N} + \frac{1}{M} \delta_f - \frac{1}{M}\delta_{f + \frac{1}{N}}$}}} \ df \nonumber\\[15pt]
&= \int\limits_{\Pi_N}[\mathcal{S}^{+}_{f,M}\mathcal{S}^{-}_{f+\frac{1}{N},M}\mathcal{L}_+(f|\xi^{M,N})P(\xi^{M,N},t)]df.
\end{aligned}
\end{equation}\end{linenomath*}
and the rate of `outflow' due to \failure is
\begin{linenomath*}\begin{equation}
\label{rate_out_lower_below}
R^{\textrm{\failure}}_{\textrm{out}}(\xi^{M,N}, t) = \underbrace{ \int\limits_{\Pi_N} }_{\substack{\text{`sum over'}\\\text{all possible $f$}}} \underbrace{\mathcal{L}_+(f | \xi^{M,N}) }_{\substack{\text{Rate of}\\ {\xi^{M,N} \to (\xi^{M,N} - \frac{1}{M}\delta_f + \frac{1}{M}\delta_{f+\frac{1}{N}})} \\ \text{transition} }}\ \underbrace{P(\xi^{M,N}, t)}_{\substack{\text{Probability of}\\\text{finding the state} \\ \text{$\xi^{M,N}$}}} \ df 
\end{equation}\end{linenomath*}
and thus the total probability flux due to \failure is
\begin{linenomath*}\begin{equation}
R^{\textrm{\failure}}_{\textrm{in}}(\xi^{M,N}, t) - R^{\textrm{\failure}}_{\textrm{out}}(\xi^{M,N}, t) = \int\limits_{\Pi_N} \left(\mathcal{S}^{+}_{f,M}\mathcal{S}^{-}_{f+\frac{1}{N},M} - 1\right)\mathcal{L}_+(f|\xi^{M,N})P(\xi^{M,N},t)df.
\end{equation}\end{linenomath*}

We omit the detailed term-by-term derivation of the rate of inflow and outflow of probability due to the other processes (\repair, disappearance, and replacement) for brevity, since they are straightforward to write down from the transition rates Eq. \ref{transition_rates} as we did above. Adding up the contributions of all possible transitions (\failure, \repair, disappearance without replacement, disappearance with replacement) to find the total probability flux through $\xi^{M,N}$ reveals that $P(\xi^{M,N}, t)$ satisfies the Master equation (Chapman-Kolmogorov equation in the maths literature):
\begin{linenomath*}\begin{align}
    \frac{\partial P}{\partial t}(\xi^{M,N},t) &= 
    \int\limits_{\Pi_N} \left(\mathcal{S}^{+}_{f,M}\mathcal{S}^{-}_{f+\frac{1}{N},M} - 1\right)\mathcal{L}_+(f|\xi^{M,N})P(\xi^{M,N},t)df\nonumber\\
    &+\int\limits_{\Pi_N} \left(\mathcal{S}^{+}_{f,M}\mathcal{S}^{-}_{f-\frac{1}{N},M} - 1\right)\mathcal{L}_-(f|\xi^{M,N})P(\xi^{M,N},t)df\nonumber\\
    &+\int\limits_{\Pi_N}\left(\mathcal{S}^{+}_{f,M} - 1\right)\mathcal{D}(f|\xi^{M,N})P(\xi^{M,N},t)df\nonumber\\
    &+ {\int\limits_{\Pi_N}\int\limits_{\Pi_N} \left(\mathcal{S}^{+}_{k,M}\mathcal{S}^{-}_{f,M}-1\right)\mathcal{R}(k,f | \xi^{M,N})P(\xi^{M,N},t)dkdf  }.
    \label{master_eqn}
\end{align}\end{linenomath*}
Each line of the RHS of Eq. \ref{master_eqn}, is, in order, the probability flux due to  \failure, \repair, disappearance without replacement, and disappearance with replacement.  For all finite $N$, the integrals in Eq. \ref{master_eqn} are to be understood as sums $\sum_{i \in \Pi_N}$, since $f$ only takes finitely many values. When $N = \infty$, they are true (Lebesgue) integrals over $[0,1]$.

\subsection{The diffusion approximation in the number of \highs ($M$)}\label{sec_supp_diffusion_approx}

We now carry out a system-size expansion (diffusion approximation) by Taylor expanding the action of the step operators in the Master Equation Eq. \ref{master_eqn}. Recall that the functional version of the Taylor expansion of a functional $F[\rho]$ about a function $\rho_0$ defined on a domain $\Omega \subseteq \mathbb{R}$ is given by
\begin{linenomath*}\begin{equation}
\label{functional_taylor_expansion}
		F[\rho_0 + \rho] = F[\rho_0] + \int\limits_{\Omega}\rho(x)\left(\frac{\delta F[\chi]}{\delta \chi(x)}\bigg{|}_{\chi = \rho_0}\right)dx + \frac{1}{2!}\int\limits_{\Omega}\int\limits_{\Omega}\rho(x)\rho(y)\left(\frac{\delta^2 F[\chi]}{\delta \chi(x)\delta \chi(y)}\bigg{|}_{\chi = \rho_0}\right)dxdy + \cdots
\end{equation}\end{linenomath*}
where $\delta F/\delta \chi$ is the functional derivative of the functional $F$, defined indirectly as the unique object that satisfies, for any smooth function $\eta$,
\begin{linenomath*}\begin{equation}
\label{functional_derivative_defn}
\int\limits_{\Omega}\frac{\delta F[\chi]}{\delta \chi(x)}\eta(x)dx = \lim_{h \to 0} \frac{F[\chi + h\eta]-F[\chi]}{h}.
\end{equation}\end{linenomath*}
The reader may easily verify that when $\rho$ and $\rho_0$ are finite-dimensional vectors (i.e. in the $N < \infty$ setting), the functional derivative Eq. \ref{functional_derivative_defn} becomes the directional derivative $\mathbf{r}\cdot\nabla F$ of a function $F:\mathbb{R}^N \to \mathbb{R}$ along a vector $\mathbf{r} \in \mathbb{R}^N$, and Eq. \ref{functional_taylor_expansion} becomes the usual multi-variate Taylor formula
\begin{linenomath*}\begin{equation}
    F(\mathbf{r_0} + \mathbf{r}) = F(\mathbf{r_0}) + \sum\limits_{i=1}^Nr_i\left(\frac{\partial F(\mathbf{x})}{\partial x_i}\bigg{|}_{\mathbf{x}=\mathbf{r_0}}\right) + \frac{1}{2!}\sum\limits_{i=1}^N\sum\limits_{j=1}^Nr_ir_j\left(\frac{\partial^2 F(\mathbf{x})}{\partial x_i\partial x_j}\bigg{|}_{\mathbf{x}=\mathbf{r_0}}\right) + \cdots 
\end{equation}\end{linenomath*}
We can now use the Taylor expansion formula Eq. \ref{functional_taylor_expansion} to expand the action of the step operators in Eq. \ref{master_eqn} in terms of $\delta_f/M$ and $(\delta_{f}-\delta_k)/M$ respectively. 

In particular, we have, for any $g, h \in \Pi_N$ and any $\mathcal{O}(M)$ functional $\mathcal{T}( \cdots | \xi)$,
\begin{linenomath*}\begin{align}
 (\mathcal{S}^{+}_{g,M} - 1 )\mathcal{T}  &=  \frac{1}{M}\int\limits_{\Pi_N}\delta_g(x)\frac{\delta \mathcal{T}}{\delta \xi(x)}dx + \frac{1}{2M^2}\int\limits_{\Pi_N}\int\limits_{\Pi_N}\delta_g(x)\delta_g(y)\frac{\delta^2 \mathcal{T}}{\delta \xi(x)\delta \xi(y)}dxdy + \mathcal{O}(M^{-2})\nonumber\\
 &= \frac{1}{M}\frac{\delta \mathcal{T}}{\delta \xi(g)} + \frac{1}{2M^2}\frac{\delta^2 \mathcal{T}}{\delta \xi(g)\delta \xi(g)} + \mathcal{O}(M^{-2})\label{singlestep_taylor}
\end{align}\end{linenomath*}
and
\begin{linenomath*}\begin{align}
 (\mathcal{S}^{+}_{g,M}\mathcal{S}^{-}_{h,M} - 1 )\mathcal{T} &=  \frac{1}{M}\int\limits_{\Pi_N}(\delta_g(x) - \delta_h(x))\frac{\delta \mathcal{T}}{\delta \xi(x)}dx\nonumber\\
 &+ \frac{1}{2M^2}\int\limits_{\Pi_N}\int\limits_{\Pi_N}(\delta_g(x) - \delta_h(x))(\delta_g(y) - \delta_h(y))\frac{\delta^2 \mathcal{T} }{\delta \xi(x)\delta \xi(y)}dxdy + \mathcal{O}(M^{-2}) \label{doublestep_taylor}
\end{align}\end{linenomath*}
Neglecting $\mathcal{O}(M^{-2})$ terms\footnote{Note that $\mathcal{T}$ is itself $\mathcal{O}(M)$ so the term $M^{-2}\int\int[\cdots]\mathcal{T}$ is $\mathcal{O}(M^{-1})$. Neglecting the higher order terms is a Gaussian approximation, justified by a functional central limit theorem such as in \citet[chapter 11]{ethier_1986_markov}. Also see \citet{pawula_approximation_1967} for why we cannot truncate a system-size expansion at any other finite order.} and Taylor expanding the RHS of Eq. \ref{master_eqn} by using:
\begin{itemize}
    \item $g = f, h = f + \frac{1}{N}, \mathcal{T} = \mathcal{L}_+(f|\xi^{M,N})P(\xi^{M,N},t)$ in Eq. \ref{doublestep_taylor} for the first line of the RHS of Eq. \ref{master_eqn}
    \item $g = f, h = f - \frac{1}{N}, \mathcal{T} = \mathcal{L}_-(f|\xi^{M,N})P(\xi^{M,N},t)$ in Eq. \ref{doublestep_taylor} for the second line of the RHS of Eq. \ref{master_eqn}
    \item $g = f, \mathcal{T} = \mathcal{D}(f | \xi^{M,N})P(\xi^{M,N},t)$ in Eq. \ref{singlestep_taylor} for the third line of the RHS of Eq. \ref{master_eqn}
    \item $ g = k, h = f, \mathcal{T} =\mathcal{R}(k,f|\xi^{M,N})P(\xi^{M,N},t)$ in Eq. \ref{doublestep_taylor} for the fourth line of the RHS of Eq. \ref{master_eqn}
\end{itemize}
yields, after some rearrangement and simplification,
\begin{linenomath*}\begin{equation}
\label{functional_FPE}
\begin{aligned}
\frac{\partial P}{\partial t}(\xi^{M,N},t) &= -\int\limits_{\Pi_N}
			\frac{\delta}{\delta\xi^{M,N}(f)}\{\mathcal{E}(f|\xi^{M,N})P(\xi^{M,N},t)\}df\\
            &+ \frac{1}{2M}\int\limits_{\Pi_N}\int\limits_{\Pi_N}\frac{\delta^2}{\delta\xi^{M,N}(f)\delta\xi^{M,N}(k)}\{\mathcal{V}(f, k|\xi^{M,N})P(\xi^{M,N},t)\}dfdk
\end{aligned}
\end{equation}\end{linenomath*}
where
\begin{linenomath*}\begin{align}
\label{FPE_drift}
    \mathcal{E}(f|\xi^{M,N}) &\coloneqq \frac{1}{M}\left[\left(\mathcal{L}_+(f-\frac{1}{N} | \xi^{M,N}) - \mathcal{L}_+(f | \xi^{M,N})\right) + \left(\mathcal{L}_-(f+\frac{1}{N} | \xi^{M,N}) - \mathcal{L}_-(f|\xi^{M,N})\right)\right]\nonumber\\
    &-\frac{1}{M}\left[\mathcal{D}(f|\xi^{M,N}) + \int\limits_{\Pi_N}\left(\mathcal{R}(f,k) - \mathcal{R}(k,f)\right)dk \right]
\end{align}\end{linenomath*}
and
\begin{linenomath*}\begin{align}
\label{FPE_diffusion}
\mathcal{V}(f, k | \xi) &\coloneqq 
\frac{\delta_k(f)}{M} \bigg[\mathcal{L}_+\left(f - \frac{1}{N} | \xi\right) +  \mathcal{L}_+(f | \xi) + \mathcal{L}_-\left(f + \frac{1}{N} | \xi\right) + \mathcal{L}_-(f | \xi)\bigg] \nonumber \\
&-\frac{1}{M} \left[ \delta_{k}\left(f+\frac{1}{N}\right)\mathcal{L}_+(f | \xi)  +\delta_{f}\left(k+\frac{1}{N}\right) \mathcal{L}_+(k | \xi) 
+ \delta_{k}\left(f-\frac{1}{N}\right)\mathcal{L}_-(f | \xi)  + \delta_{f}\left(k-\frac{1}{N}\right)\mathcal{L}_-(k | \xi) \right]\nonumber\\
&+\frac{\delta_k(f)}{M} \bigg[\mathcal{D}(f | \xi) + \int_{\Pi_N} (\mathcal{R}(f, y | \xi) +  \mathcal{R}(y, f | \xi) )dy \bigg] \nonumber \\
&-\frac{1}{M} \left[ \mathcal{R}(f, k | \xi) + \mathcal{R}(k, f | \xi) \right].
\end{align}\end{linenomath*}
The first and third lines on the RHS of Eq. \ref{FPE_diffusion} are the `diagonal' terms and represent stochasticity at the point $f$ arising from the events that occur at that point (due to \failure/\repair in the first line and disappearance/replacement on the third). The second and fourth lines collect the `off-diagonal' terms and account for the correlations in fluctuations between states due to both lower level dynamics (the total number of \lows per \high is strictly constant, accounted for in the second line\footnote{note that we have moved the $\pm 1/N$ shifts of the point masses to the arguments rather than the locations of the point masses to make the presentation cleaner. For instance, $\int[\cdots]\delta_f(k+\frac{1}{N})dk$ is equivalent to $\int[\cdots]\delta_{f-\frac{1}{N}}(k)dk$.} of the RHS of Eq. \ref{FPE_diffusion}), and replacement at the higher level (one \high disappearing instantaneously being compensated for via replacement by a copy of another \high, accounted for by the fourth line of the RHS of Eq. \ref{FPE_diffusion}). Equation \ref{functional_FPE} is a (functional) Fokker-Planck equation (Kolmogorov forward equation in the maths literature). A process whose probability density satisfies Eq. \ref{functional_FPE} must solve the It\^o stochastic (partial) differential equation~\citep{gardiner_stochastic_2009}
\begin{linenomath*}\begin{equation}
\label{supp_SPDE}
    \frac{\partial \xi^{M,N}}{\partial t}(f,t) = \mathcal{E}(f,\xi^{M,N}) + \frac{1}{\sqrt{M}} \eta_{\xi^{M,N}}(f, t)
\end{equation}\end{linenomath*}
where $\eta_{\xi^{M,N}}(f,t)$ is a Gaussian spacetime noise process or Gaussian random field~\citep[chapter 4]{daprato_spde_2014} that is white in time and coloured in space according to the correlations induced by Eq. \ref{FPE_diffusion}. In particular, given any square integrable function $G$ on $\Pi_N \times [0,\infty)$ and any time $t > 0$,  the process $\eta_{\xi^{M,N}}$ obeys
\begin{linenomath*}\begin{equation}
\label{spde_noise_mean}
\mathbb{E}\left[\int\limits_{0}^{t} \int\limits_{\Pi_N} G(u, s) \eta_{\xi^{M,N}}(u, s) du ds \right] = 0,
\end{equation}\end{linenomath*}
and given any two such square integrable functions $G, H$ and any two times $t_1, t_2 > 0$, it obeys
\begin{linenomath*}\begin{equation}
\label{spde_noise_var}
\begin{aligned}
&\mathbb{E}\left[\int\limits_{0}^{t_1} \int\limits_{\Pi_N} G(u, s) \eta_{\xi^{M,N}}(u, s) du ds \int\limits_{0}^{t_2} \int\limits_{\Pi_N} H(v, r) \eta_{\xi^{M,N}}(v, r) dv dr \right] \\[12pt] &=\int\limits_{0}^{\min(t_1,t_2)} \int\limits_{\Pi_N} \int\limits_{\Pi_N} G(u, s) H(v, s) \mathcal{V}(u, v | \xi^{M,N}) du dv ds.
\end{aligned}
\end{equation}\end{linenomath*}
Substituting the functional forms of the infinitesimal rates $\mathcal{L}_\pm, \mathcal{D}$, and $\mathcal{R}$ from Eq. \ref{transition_functional_forms} into the $\mathcal{E}$ term in Eq. \ref{supp_SPDE} yields our main equation, Eq. \ref{spde_maintext}. The correlation structure of the noise term is defined by Eq. \ref{spde_noise_mean} and Eq. \ref{spde_noise_var}. As an aside, we note that the martingale representation theorem (see e.g.  \citet{daprato_spde_2014}, Theorem 8.2) guarantees the existence of a function $\mathcal{S}(f,g | \xi^{M,N})$ defined in terms of $\mathcal{V}$ (analogous to ``taking a square root'') such that
\begin{linenomath*}\begin{equation*}
    \eta_{\xi^{M,N}}(f,t) = \int\limits_{\Pi_N}\mathcal{S}(f,k | \xi^{M,N})\dot{W}(k,t)dk
\end{equation*}\end{linenomath*}
where $\dot{W}$ is a spacetime white noise on $\Pi_N \times [0,\infty)$.

\section{Deriving an SPDE for the limiting superprocess when  $\mu = N\mu_e + \nu(f)$ and $N\to\infty, M \to \infty$ together such that $N/M = \lambda$}\label{supp_sec_superprocess}

Since we have a Fokker-Planck equation Eq. \ref{functional_FPE}, we can take the relevant limit in $\mathcal{E}$ and $\mathcal{V}$ from Eq. \ref{FPE_drift} and Eq. \ref{FPE_diffusion} to derive the resultant drift (expected behaviour) and diffusion (covariance structure) for $\zeta$. In other words, by taking our relevant limit on both sides of Eq. \ref{functional_FPE}, we arrive at a functional Fokker-Planck equation
\begin{linenomath*}\begin{equation}
\label{superprocess_FPE}
\begin{aligned}
\frac{\partial P}{\partial t}(\zeta^\lambda,t) &= -\int\limits_{[0,1]}
			\frac{\delta}{\delta\zeta^\lambda(f)}\{\mathcal{E}_{\zeta^\lambda}(f|\zeta^\lambda)P(\zeta^\lambda,t)\}df\\
            &+ \frac{1}{2}\int\limits_{[0,1]}\int\limits_{[0,1]}\frac{\delta^2}{\delta\zeta^\lambda(f)\delta\zeta^\lambda(k)}\{\mathcal{V}_{\zeta^\lambda}(f, k|\zeta^\lambda)P(\zeta^\lambda,t)\}dfdk
\end{aligned}
\end{equation}\end{linenomath*}
where
\begin{linenomath*}\begin{align}
   \mathcal{E}_{\zeta^\lambda}(f|\zeta^\lambda) &=  \lim\limits_{\substack{N\to\infty\\M\to\infty\\\frac{N}{M} = \lambda}} \mathcal{E}(f|\xi^{M,N}) \quad ,\label{superprocess_drift_defn}\\
   \mathcal{V}_{\zeta^\lambda}(f, k|\zeta^\lambda) &=  \lim\limits_{\substack{N\to\infty\\M\to\infty\\\frac{N}{M} = \lambda}} \frac{1}{M} \mathcal{V}(f, k|\xi^{M,N}).\label{superprocess_diffusion_defn}
\end{align}\end{linenomath*}
are to be found and will determine the SPDE for $\zeta^\lambda$.

\subsection{The expected behaviour of the limiting superprocess}\label{sec_superprocess_mean}

We start with $\mathcal{E}_{\zeta^\lambda}$. Directly substituting the functional form of $\mathcal{L}_\pm$ from Eq. \ref{transition_functional_forms} into Eq. \ref{FPE_drift} gives
\begin{linenomath*}\begin{equation}
\resizebox{\textwidth}{!}{%
$\displaystyle
\mathcal{L}_\mp(f\pm\frac{1}{N}) - \mathcal{L}_\mp(f) = MN\left(r_{\mp}(f\pm\frac{1}{N})\xi^{M,N}(f\pm\frac{1}{N}) - r_\mp(f)\xi^{M,N}(f)\right) = MN\left(\pm\frac{1}{N}\frac{\partial}{\partial f}\{ r_{\mp}(f)\xi^{M,N}(f)\} + \mathcal{O}(N^{-2})\right)
$
}
\end{equation}\end{linenomath*}
and thus
\begin{linenomath*}\begin{equation}
\label{superprocess_drift_L}
\lim\limits_{\substack{N\to\infty\\M\to\infty\\\frac{N}{M}=\lambda}} \frac{1}{M}\left( \mathcal{L}_\mp(f\pm\frac{1}{N}) - \mathcal{L}_\mp(f)\right) = \lim\limits_{\substack{N\to\infty\\M\to\infty\\\frac{N}{M}=\lambda}}\left(\pm\frac{\partial}{\partial f}\{ r_{\mp}(f)\xi^{M,N}(f)\} + \mathcal{O}(N^{-1})\right) = \pm\frac{\partial}{\partial f}\{ r_{\mp}\zeta^{\lambda}(f)\}.
\end{equation}\end{linenomath*}
After substituting $\repprob = 1 - (\varepsilon/M)$, the contribution of disappearance and replacement to $\mathcal{E}$ is
\begin{linenomath*}\begin{equation}
\label{superprocess_drift_mort}
    -\frac{1}{M}\left[\mathcal{D}(f|\xi^{M,N}) + \int\limits_{\Pi_N}\left(\mathcal{R}(f,k) - \mathcal{R}(k,f)\right)dk \right] = - \left( \frac{\varepsilon}{M}\mu(f) +  \left(1-\frac{\varepsilon}{M}\right)\left[\mu(f)-{\widehat{\mu}(t)}\right] \right) \ \xi^{M,N}(f).
\end{equation}\end{linenomath*}
where $\widehat{\mu}(t) \coloneqq \overline{\mu}(t)/\ell(t)$.
Plugging in the RHS of Eq. \ref{superprocess_drift_L} and Eq. \ref{superprocess_drift_mort} into Eq. \ref{superprocess_drift_defn}, substituting $\mu(f) = N\mu_e + \nu(f)$, and taking the limit reveals that
\begin{linenomath*}\begin{equation}
    \mathcal{E}_{\zeta^\lambda}(f|\zeta^\lambda) = -\frac{\partial}{\partial f}\left\{r(f)\zeta^\lambda(f)\right\} - \left[\varepsilon\lambda\mu_e +\nu(f) - {\widehat{\nu}(t)}\right]\zeta^{\lambda}(f) 
\end{equation}\end{linenomath*}
where $r(f) \coloneqq r_+(f) - r_-(f)$ as in the rest of the manuscript and
\begin{linenomath*}\begin{equation*}
    \widehat{\nu}(t) \coloneqq \frac{\displaystyle\int\limits_0^1\nu(f)\zeta^\lambda(f,t)df}{\displaystyle\int\limits_0^1\zeta^\lambda(f,t)df}.
\end{equation*}\end{linenomath*}

\subsection{The covariance structure of the limiting superprocess}\label{sec_superprocess_var}

We now move to  $\mathcal{V}_{\zeta^{\lambda}}$. By examining Eq. \ref{FPE_diffusion}, we see that we can decompose $\mathcal{V}_{\zeta^{\lambda}}$ into the sum of two terms, $\mathcal{V}_{\zeta^{\lambda}} = \mathcal{V}^{\text{high}}_{\zeta^\lambda} + \mathcal{V}^{\text{low}}_{\zeta^\lambda}$, where $\mathcal{V}^{\text{high}}_{\zeta^\lambda}$ collects terms involving the higher level dynamics (disappearance and replacement) and $\mathcal{V}^{\text{low}}_{\zeta^\lambda}$ collects terms involving the lower level dynamics (\failure and \repair).

For the higher-level contributions, we can further distinguish between the terms coming from disappearance without replacement, and those coming from disappearance with replacement. The contributions to $\mathcal{V}^{\text{high}}_{\zeta^\lambda}$ from $\mathcal{D}$ are (from Eq. \ref{FPE_diffusion} and Eq. \ref{superprocess_diffusion_defn}, all of the form
\begin{linenomath*}\begin{equation}
  \frac{1}{M^2}\mathcal{D}(f|\xi) = \frac{1}{M}\left(\frac{\varepsilon}{M}(N\mu_e + \nu(f))\xi^{M,N}\right)   = \frac{1}{M}\left(\varepsilon\lambda\mu_e + \frac{\varepsilon}{M}\nu(f))\xi^{M,N}\right) 
\end{equation}\end{linenomath*}
and thus vanish as $M \to \infty$. Disappearance without replacement thus does not contribute to the infinitesimal variance of our limiting process. 

The contributions from the replacement dynamics are terms the form $\mathcal{R}(g,h) + \mathcal{R}(h,g)$, multiplied by $M^{-2}$ (there is an $M^{-1}$ in the definition of $\mathcal{V}$ in Eq. \ref{FPE_diffusion}, and an additional $M^{-1}$ in Eq. \ref{superprocess_diffusion_defn}). Substituting $\repprob = 1 - (\varepsilon/M)$ and $\mu = N\mu_e + \nu(f)$ gives
\begin{linenomath*}\begin{equation}
\label{int_for_high_diffusion}
      \frac{1}{M^2}\left(
      \mathcal{R}(g,h) + \mathcal{R}(h,g)\right)= \frac{1}{\ell(t)}\left(1-\frac{\varepsilon}{M}\right)\left(2\lambda\mu_e + \frac{1}{M}\left[\nu(g) + \nu(h)\right]\right)\xi^{M,N}(g)\xi^{M,N}(h) 
\end{equation}\end{linenomath*}
Upon taking the limits $N \to \infty, M \to \infty$ with $N/M$ held at $\lambda$, the terms with $1/M$ vanish and only the $2\lambda\mu_e$ (along with the product of densities) remains. Substituting into Eq. \ref{superprocess_diffusion_defn}, taking the limit, and rearranging slightly, we see that the entire $\mathcal{V}^{\text{high}}_{\zeta^\lambda}$ term in the limit is given by
\begin{linenomath*}\begin{equation}
\label{superprocess_high_diffusion}
    \mathcal{V}^{\text{high}}_{\zeta^\lambda} = 2\lambda\mu_e\bigg[\delta_k(f)\zeta(f)-\frac{1}{\ell(t)}\zeta(k)\zeta(f)\bigg]
\end{equation}\end{linenomath*}
We will now show that $\mathcal{V}^{\text{low}}_{\zeta^\lambda} = 0$ in the limit.

\subsubsection{Lower-level dynamics do not contribute to noise in the stochastic limit}

For brevity, we separate the contributions of $\mathcal{L}_+$ and $\mathcal{L}_-$ by  writing $\mathcal{V}^{\text{low}}_{\zeta^\lambda} = \mathcal{V}^{\mathcal{L}_+}_{\zeta^\lambda} + \mathcal{V}^{\mathcal{L}_-}_{\zeta^\lambda} $ and demonstrate here that $\mathcal{V}^{\mathcal{L}_+}_{\zeta^\lambda}$ vanishes in the limit. The argument for $\mathcal{V}^{\mathcal{L}_-}_{\zeta^\lambda}$ follows \emph{mutatis mutandis} because all terms that occur are entirely symmetric.  Collecting all the $\mathcal{L}_{+}$ terms in Eq. \ref{FPE_diffusion} and absorbing the additional $M^{-1}$ from the RHS of Eq. \ref{superprocess_diffusion_defn}, we require the limit of
\begin{linenomath*}\begin{equation}
\label{int_for_superdiffusion}
    \mathcal{V}^{\mathcal{L}_+}  \coloneqq  M^{-2}\left[
\delta_k(f)\mathcal{L}_+\left(f - \frac{1}{N} | \xi\right) +  \delta_k(f)\mathcal{L}_+(f | \xi)
- \delta_{k}\left(f+\frac{1}{N}\right)\mathcal{L}_+(f | \xi) - \delta_{f}\left(k+\frac{1}{N}\right) \mathcal{L}_+(k | \xi)  \right]
\end{equation}\end{linenomath*}
We now employ some mathematical trickery. Since the entire $\mathcal{V}$ term (and thus the limit we are interested in) only appears within double integrals in equation \ref{functional_FPE} and Eq. \ref{superprocess_FPE}, products of functionals with point masses obey 
$\int\int\mathcal{F}(y)\delta_x(y+a)dxdy = \int \int\mathcal{F}(x-a)\delta_x(y)dxdy =  \int \int\mathcal{F}(x)\delta_{x-a}(y)dxdy =  \int \int\mathcal{F}(x)\delta_{y+a}(x)dxdy$.
In other words, we can shift the argument of the functional being integrated and move the absolute location of the point mass without affecting the final expression as long as their relative positions are respected. 
We thus have $\delta_{f}\left(k+\frac{1}{N}\right) \mathcal{L}_+(k | \xi) = \delta_{k}\left(f-\frac{1}{N}\right) \mathcal{L}_+(f - \frac{1}{N} | \xi)$. Substituting into Eq. \ref{int_for_superdiffusion} and rearranging slightly gives
\begin{linenomath*}\begin{equation}
    \mathcal{V}^{\mathcal{L}_+}  = - M^{-2}\left[
 \left(\delta_{k}(f+\frac{1}{N}) - \delta_k(f) \right)\mathcal{L}_+(f | \xi) 
 + \left(\delta_{k}(f-\frac{1}{N}) - \delta_k(f)\right) \mathcal{L}_+(f - \frac{1}{N} | \xi) \right]
\end{equation}\end{linenomath*}
Adding and subtracting $\left(\delta_{k}\left(f-\frac{1}{N}\right) - \delta_k(f) \right)\mathcal{L}_+(f | \xi)$ within the square brackets on the RHS yields
\begin{linenomath*}\begin{equation}
\label{int_2_for_superdiffusion}
\begin{aligned}
    \mathcal{V}^{\mathcal{L}_+}  &= - M^{-2}
 \left(\delta_{k}(f+\frac{1}{N})  - 2\delta_k(f)+\delta_{k}(f-\frac{1}{N})\right)\mathcal{L}_+(f | \xi)\\ 
 &\hphantom{=}- M^{-2}\left(\delta_{k}(f-\frac{1}{N}) - \delta_k(f)\right) \left(\mathcal{L}_+(f - \frac{1}{N} | \xi) - \mathcal{L}_+(f | \xi) \right)
 \end{aligned}
\end{equation}\end{linenomath*}
Now substituting the functional form of $\mathcal{L}_+$, we note that it contains $MN$, which together with the $M^{-2}$ outside the brackets in Eq. \ref{int_2_for_superdiffusion} gives $N/M = \lambda$. Letting $R_+(f) \coloneqq r_+(f)\xi^{M,N}(f,t)$ for concision, we thus have
\begin{linenomath*}\begin{equation}
\label{int_3_for_superdiffusion}
\begin{aligned}
    \mathcal{V}^{\mathcal{L}_+}  &= - \lambda
 \left(\delta_{k}(f+\frac{1}{N})  - 2\delta_k(f)+\delta_{k}(f-\frac{1}{N})\right)R_+(f)\\ 
 &\hphantom{=}- \lambda\left(\delta_{k}(f-\frac{1}{N}) - \delta_k(f)\right) \left(R_+(f - \frac{1}{N}) - R_+(f) \right)
 \end{aligned}
\end{equation}\end{linenomath*}
Let $h \coloneqq 1/N$. Multiplying and dividing the RHS of Eq. \ref{int_3_for_superdiffusion} by $h^2$ gives us
\begin{linenomath*}\begin{equation}
\label{int_4_for_superdiffusion}
\begin{aligned}
    \mathcal{V}^{\mathcal{L}_+}  &= - \lambda h^2
 \left(\frac{\delta_{k}(f+h)  - 2\delta_k(f)+\delta_{k}(f-h)}{h^2}\right) R_+(f)\\ 
 &\hphantom{=}- \lambda h^2 \left(\frac{\delta_{k}(f-h) - \delta_k(f)}{h}\right) \left(\frac{R_+(f - h) - R_+(f)}{h} \right)
 \end{aligned}
\end{equation}\end{linenomath*}
The fraction within the parentheses on the first line of the RHS of Eq. \ref{int_4_for_superdiffusion} is a second-order difference quotient. The two fractions on the second line of the RHS are first-order difference quotients. As we take our limit, $h \to 0$ (because $N \to \infty$) and these quotients converge to distributional derivatives (see e.g. exercise 22 in chapter 6 of \citet{rudin_functional_1991}). In other words, as the `granularity' of our process becomes finer ($N \to \infty$), the changes due to the `local' jumps between adjacent urns, initially described by finite/discrete difference quotients, end up being described by derivatives
\begin{linenomath*}\begin{align*}
   \frac{\delta_{k}(f+h)  - 2\delta_k(f)+\delta_{k}(f-h)}{h^2} \ &\to \ \hphantom{-}\frac{\partial^2 \delta_k(f)}{\partial f^2}  \\
  \frac{\delta_{k}(f-h) - \delta_k(f)}{h} \ &\to  \ -\frac{\partial \delta_k(f)}{\partial f} \ , \quad \text{and}\\
\frac{R_+(f - h) - R_+(f)}{h} \ &\to  \ -\frac{\partial R_+(f)}{\partial f}
\end{align*}\end{linenomath*}
Distributional derivatives like $\partial\delta_k/\partial f$ are formally defined in the weak sense~\citep[chapter 6]{rudin_functional_1991} following a generalisation of the usual integration-by-parts formula. For our purposes, the important bit is that this definition ensures distributional derivatives are finite when (doubly) integrated (alongside test functions). Thus, the left over pre-factors of $h^2$ on the RHS of Eq. \ref{int_4_for_superdiffusion} ensure that
\begin{linenomath*}\begin{align*}
   h^2 \int_{\Pi_N} \int_{\Pi_N}  [\cdots]  \frac{\partial^2 \delta_k(f)}{\partial f^2} dfdk \ \to \ 0 & \ \text{ as  }  N\to \infty, \quad \text{and}\\
   h^2 \int_{\Pi_N} \int_{\Pi_N}  [\cdots]\frac{\partial \delta_k(f)}{\partial f}\frac{\partial R_+(f)}{\partial f} df dk \ \to  \ 0 & \ \text{ as  }  N\to \infty.\\
\end{align*}\end{linenomath*}
This demonstrates that
\begin{linenomath*}\begin{equation}
    \mathcal{V}^{\mathcal{L}_+}_{\zeta^\lambda} \coloneqq  \lim\limits_{\substack{N\to\infty\\M\to\infty\\\frac{N}{M}=\lambda}}  \mathcal{V}^{\mathcal{L}_+} = 0.
\end{equation}\end{linenomath*}
An identical argument shows that $\mathcal{V}^{\mathcal{L}_-}_{\zeta^\lambda} = 0$ as well, and thus $\mathcal{V}^{\text{low}}_{\zeta^\lambda} = 0$. In other words, stochasticity due to the lower-level dynamics of \failure and \repair vanish in this limit.

We have hence shown that $\mathcal{V}_{\zeta^\lambda}$ in Eq. \ref{superprocess_FPE} is given only by Eq. \ref{superprocess_high_diffusion}, i.e.
\begin{linenomath*}\begin{equation}
\label{superprocess_diffusion}
    \mathcal{V}_{\zeta^\lambda}(f, k|\zeta^\lambda) = 2\lambda\mu_e\bigg[\delta_k(f)\zeta(f)-\frac{1}{\ell(t)}\zeta(k)\zeta(f)\bigg]
\end{equation}\end{linenomath*}
Notice that when $\ell(t) = 1$, \emph{i.e.} replacement is guaranteed, we exactly recover the covariance (more properly, the quadratic variation) of a (scaled) Fleming-Viot process~\citep[section 4.2]{etheridge_2000_superprocesses}. When $\repprob < 1$ (i.e. $\varepsilon > 0$  and $\ell(t) < 1$), our process is not a Fleming-Viot process proper because it is not probability-measure-valued (it is sub-probability-measure-valued). Since $\lambda\mu_e$ is a constant that scales the strength of the noise, it is useful to define
\begin{linenomath*}\begin{equation}
\label{superprocess_variance_diffusion_rescaled}
\mathcal{V}^{\text{FV-like}}_{\zeta^\lambda}(f, k|\zeta^\lambda) \coloneqq 2\bigg[\delta_k(f)\zeta(f)-\frac{1}{\ell(t)}\zeta(k)\zeta(f)\bigg]
\end{equation}\end{linenomath*}
so that Eq. \ref{superprocess_diffusion}  can be written as $\mathcal{V}_{\zeta^\lambda}(f, k|\zeta^\lambda) = \lambda\mu_e \mathcal{V}^{\text{FV-like}}_{\zeta^\lambda}(f, k|\zeta^\lambda)$.

\subsection{Putting it all together}

Putting the results of section \ref{sec_superprocess_mean} and section \ref{sec_superprocess_var} together and translating from the functional Fokker-Planck equation Eq. \ref{superprocess_FPE} to the equivalent SPDE representation, we have thus shown that the limiting process $\zeta^{\lambda}$ satisfies the SPDE
\begin{linenomath*}\begin{equation}
\label{supp_superprocess_spde}
    \frac{\partial \zeta^{\lambda}}{\partial t}(f,t) = \mathcal{E}_{\lambda}[f,\zeta^{\lambda}] +  \sqrt{\mu_e\lambda} \ \eta_{\zeta^{\lambda}}(f,t)
\end{equation}\end{linenomath*}
where
\begin{linenomath*}\begin{equation}   
\label{supp_superprocess_spde_mean}
    \mathcal{E}_{\lambda}[f,\zeta^{\lambda}] = -\frac{\partial}{\partial f}\left\{r(f)\zeta^\lambda(f,t)\right\} - \left[\varepsilon\lambda\mu_e + \bigg(\nu(f)-\widehat{\nu}(t)\bigg)\right]\zeta^{\lambda}(f,t) 
\end{equation}\end{linenomath*}
and $\eta_{\zeta^\lambda}$ is a noise process (random field) such that, given any square integrable function $G$ on $[0,1] \times [0,\infty)$ and any time $t > 0$,  the process $\eta_{\zeta^\lambda}$ obeys
\begin{linenomath*}\begin{equation}
\label{superprocess_spde_noise_mean}
\mathbb{E}\left[\int\limits_{0}^{t} \int\limits_{0}^{1} G(u, s) \eta_{\zeta^\lambda}(u, s) du ds \right] = 0,
\end{equation}\end{linenomath*}
and given any two such square integrable functions $G, H$ and any two times $t_1, t_2 > 0$, it obeys
\begin{linenomath*}\begin{equation}
\label{superprocess_spde_noise_var}
\begin{aligned}
&\mathbb{E}\left[\int\limits_{0}^{t_1} \int\limits_{0}^{1} G(u, s)\eta_{\zeta^\lambda}(u, s) du ds \int\limits_{0}^{t_2} \int\limits_{0}^{1} H(v, r) \eta_{\zeta^\lambda}(v, r) dv dr \right] \\[12pt] &=\int\limits_{0}^{\min(t_1,t_2)} \int\limits_{0}^{1} \int\limits_{0}^{1} G(u, s) H(v, s) \mathcal{V}^{\text{FV-like}}_{\zeta^\lambda}(u, v | \zeta^\lambda) du dv ds,
\end{aligned}
\end{equation}\end{linenomath*}
where 
\begin{linenomath*}\begin{equation}
\mathcal{V}^{\text{FV-like}}_{\zeta^\lambda}(u, v|\zeta^\lambda) \coloneqq 2\bigg[\delta_v(u)\zeta(u)-\frac{1}{\ell(t)}\zeta(u)\zeta(v)\bigg].
\end{equation}\end{linenomath*}
The SPDE \ref{supp_superprocess_spde} is Eq. \ref{superprocess_spde} in the main text.

\begin{refcontext}[sorting=nyc]
\printbibliography[title=References Cited in the Supplement]\addcontentsline{toc}{section}{References Cited in the Supplement}

@article{moorad_2019_extrinsic,
  title = {Evolutionary Ecology of Senescence and a Reassessment of Williams’ ‘Extrinsic Mortality’ Hypothesis},
  volume = {34},
  DOI = {10.1016/j.tree.2019.02.006},
  number = {6},
  ISSN = {0169-5347},
  journal = {Trends in Ecology \& Evolution},
  author = {Moorad,  Jacob and Promislow,  Daniel and Silvertown,  Jonathan},
  year = {2019},
  pages = {519–530},
}

@book{fischer_four_2024,
  title={Four views on free will},
  author={Fischer, John Martin and Kane, Robert and Pereboom, Derk and Vargas, Manuel},
  year={2024},
  publisher={Wiley-Blackwell},
  isbn={978-1-394-16196-6},
  edition={2},
}

@article{dennett_2001_hunch,
  title = {The Zombic Hunch: Extinction of an Intuition?},
  volume = {48},
  DOI = {10.1017/s1358246100010687},
  journal = {Royal Institute of Philosophy Supplement},
  publisher = {Cambridge University Press (CUP)},
  author = {Dennett,  Daniel},
  year = {2001},
  pages = {27–43}
}

@article{agren_agency_2022,
  title = {Genetic conflicts and the case for licensed anthropomorphizing},
  volume = {76},
  ISSN = {1432-0762},
  url = {http://dx.doi.org/10.1007/s00265-022-03267-6},
  DOI = {10.1007/s00265-022-03267-6},
  number = {12},
  journal = {Behavioral Ecology and Sociobiology},
  publisher = {Springer Science and Business Media LLC},
  author = {\AA gren,  J. Arvid and Patten,  Manus M.},
  year = {2022},
  month = Dec 
}

@unpublished{bhat_failures_2026,
  title = {Stochastic failure accumulation as a foundation for exponential mortality and selective disappearance},
  DOI = {10.64898/2026.05.25.727614},
  note = {Available on BioRxiv},
  author = {Bhat,  Ananda Shikhara and Kokko,  Hanna},
  year = {submitted},
}

@article{cooney_2025_cultural,
  title = {Exploring the Evolution of Altruistic Punishment with a PDE Model of Cultural Multilevel Selection},
  volume = {87},
  ISSN = {1522-9602},
  url = {http://dx.doi.org/10.1007/s11538-025-01422-4},
  DOI = {10.1007/s11538-025-01422-4},
  number = {4},
  journal = {Bulletin of Mathematical Biology},
  publisher = {Springer Science and Business Media LLC},
  author = {Cooney,  Daniel B.},
  year = {2025},
  month = Mar 
}

@article{week_gmatrix_2026,
  title = {Stochastic eco-evolutionary dynamics of multivariate traits: A framework for modeling population processes illustrated by the study of drifting G-matrices},
  volume = {625},
  ISSN = {0022-5193},
  DOI = {10.1016/j.jtbi.2026.112428},
  journal = {Journal of Theoretical Biology},
  author = {Week,  Bob},
  year = {2026},
  month = May,
  pages = {112428}
}

@inbook{birkner_FV_2009,
  title = {Measure-valued diffusions,  coalescents and genetic inference},
  ISBN = {9781139107020},
  url = {http://dx.doi.org/10.1017/CBO9781139107020.015},
  DOI = {10.1017/cbo9781139107020.015},
  booktitle = {Trends in Stochastic Analysis},
  publisher = {Cambridge University Press},
  author = {Birkner,  Matthias and Blath,  Jochen},
  year = {2009},
  month = Apr,
  pages = {329–364}
}

@article{rossine_2025_plasmid,
  title = {Intracellular competition shapes plasmid population dynamics},
  volume = {390},
  ISSN = {1095-9203},
  url = {http://dx.doi.org/10.1126/science.adx0665},
  DOI = {10.1126/science.adx0665},
  number = {6779},
  journal = {Science},
  publisher = {American Association for the Advancement of Science (AAAS)},
  author = {Rossine,  Fernando and Sanchez,  Carlos and Eaton,  Daniel and Paulsson,  Johan and Baym,  Michael},
  year = {2025},
  month = Dec 
}

@article{paulsson_2002_plasmid,
  title = {Multileveled Selection on Plasmid Replication},
  volume = {161},
  ISSN = {1943-2631},
  DOI = {10.1093/genetics/161.4.1373},
  number = {4},
  journal = {Genetics},
  author = {Paulsson,  Johan},
  year = {2002},
  pages = {1373–1384}
}

@article{williams_1999_tithonus,
  title = {The Tithonus Error in Modern Gerontology},
  volume = {74},
  ISSN = {1539-7718},
  url = {http://dx.doi.org/10.1086/394111},
  DOI = {10.1086/394111},
  number = {4},
  journal = {The Quarterly Review of Biology},
  publisher = {University of Chicago Press},
  author = {Williams,  George C.},
  year = {1999},
  month = Dec,
  pages = {405–415}
}

@article{fleming_process_1979,
 author = {Fleming, Wendell H. and Viot, Michel},
 journal = {Indiana University Mathematics Journal},
 number = {5},
 pages = {817--843},
 publisher = {Indiana University Mathematics Department},
 title = {Some Measure-Valued Markov Processes in Population Genetics Theory},
 volume = {28},
 year = {1979}
}

@book{okasha_2018_agents,
  title = {Agents and Goals in Evolution},
  ISBN = {9780198815082},
  DOI = {10.1093/oso/9780198815082.001.0001},
  publisher = {Oxford University Press},
  author = {Okasha, Samir},
  year = {2018},
}

@article{mcleod_social_2019,
	title = {Social evolution under demographic stochasticity},
	volume = {15},
	issn = {1553-7358},
	doi = {10.1371/journal.pcbi.1006739},
	number = {2},
	journal = {PLOS Computational Biology},
	author = {McLeod, David V. and Day, Troy},
	year = {2019},
	pages = {e1006739},
}

@article{raatz_promoting_2023,
	title = {Promoting extinction or minimizing growth? {The} impact of treatment on trait trajectories in evolving populations},
	volume = {77},
	issn = {0014-3820},
	doi = {10.1093/evolut/qpad042},
	number = {6},
	journal = {Evolution},
	author = {Raatz, Michael and Traulsen, Arne},
	month = jun,
	year = {2023},
	pages = {1408--1421},
}

@book{daprato_spde_2014,
  title = {Stochastic Equations in Infinite Dimensions},
  ISBN = {9781107295513},
  DOI = {10.1017/cbo9781107295513},
  publisher = {Cambridge University Press},
  author = {Da Prato,  Giuseppe and Zabczyk,  Jerzy},
  year = {2014},
  series = {Encyclopedia of Mathematics and its Applications},
  number = 152,
}

@book{rudin_functional_1991,
  title={Functional Analysis},
  author={Rudin, W.},
  isbn={978-0070619883},
  series={International Series in Pure and Applied Mathematics},
  year={1991},
  publisher={McGraw-Hill},
  address={New York},
  edition = {2}
}

@article{parsons_2017_manifold,
  title = {Dimension reduction for stochastic dynamical systems forced onto a manifold by large drift: a constructive approach with examples from theoretical biology},
  volume = {50},
  ISSN = {1751-8121},
  url = {http://dx.doi.org/10.1088/1751-8121/aa86c7},
  DOI = {10.1088/1751-8121/aa86c7},
  number = {41},
  journal = {Journal of Physics A: Mathematical and Theoretical},
  publisher = {IOP Publishing},
  author = {Parsons,  Todd L and Rogers,  Tim},
  year = {2017},
  pages = {415601},
}

@book{garcia-ojalvo_noise_1999,
	address = {New York},
	series = {Institute for nonlinear science ({Springer}-{Verlag})},
	title = {Noise in spatially extended systems},
	isbn = {978-0-387-98855-9},
	urldate = {2024-05-28},
	publisher = {Springer},
	author = {García-Ojalvo, Jordi and Sancho, Jose M.},
	year = {1999},
}

@article{constable_2014_fast,
  title = {Fast-mode elimination in stochastic metapopulation models},
  volume = {89},
  ISSN = {1550-2376},
  url = {http://dx.doi.org/10.1103/PhysRevE.89.032141},
  DOI = {10.1103/physreve.89.032141},
  number = {3},
  journal = {Physical Review E},
  publisher = {American Physical Society (APS)},
  author = {Constable,  George W. A. and McKane,  Alan J.},
  year = {2014},
}

@article{constable_2013_manifold,
  title = {Stochastic dynamics on slow manifolds},
  volume = {46},
  ISSN = {1751-8121},
  url = {http://dx.doi.org/10.1088/1751-8113/46/29/295002},
  DOI = {10.1088/1751-8113/46/29/295002},
  number = {29},
  journal = {Journal of Physics A: Mathematical and Theoretical},
  publisher = {IOP Publishing},
  author = {Constable,  George W A and McKane,  Alan J and Rogers,  Tim},
  year = {2013},
  pages = {295002}
}

@article{mgonigle_2026_prudence,
  title = {Structured Landscapes Promote Persistence by Favoring Prudent Predators},
  volume = {208},
  ISSN = {1537-5323},
  url = {http://dx.doi.org/10.1086/740827},
  DOI = {10.1086/740827},
  number = {1},
  journal = {The American Naturalist},
  publisher = {University of Chicago Press},
  author = {M’Gonigle,  Leithen K. and Green,  Emma L. and Greenspoon,  Philip B.},
  year = {2026},
  pages = {72–82},
}

@book{etheridge_2000_superprocesses,
  title = {An Introduction to Superprocesses},
  ISBN = {9780821827062},
  DOI = {10.1090/ulect/020},
  series = {University Lecture Series},
  publisher = {American Mathematical Society},
  author = {Etheridge,  Alison},
  year = {2000}}

@article{weitz_2001_explaining,
  title = {Explaining mortality rate plateaus},
  volume = {98},
  ISSN = {1091-6490},
  DOI = {10.1073/pnas.261228098},
  number = {26},
  journal = {Proceedings of the National Academy of Sciences},
  publisher = {Proceedings of the National Academy of Sciences},
  author = {Weitz,  Joshua S. and Fraser,  Hunter B.},
  year = {2001},
  month = dec,
  pages = {15383–15386}
}

@article{kerr_2006_prudent,
  title = {Local migration promotes competitive restraint in a host–pathogen “tragedy of the commons”},
  volume = {442},
  DOI = {10.1038/nature04864},
  number = {7098},
  journal = {Nature},
  author = {Kerr,  Benjamin and Neuhauser,  Claudia and Bohannan,  Brendan J. M. and Dean,  Antony M.},
  year = {2006},
  pages = {75–78}
}

@article{rankin_2007_tragedy,
  title = {The tragedy of the commons in evolutionary biology},
  volume = {22},
  ISSN = {0169-5347},
  url = {http://dx.doi.org/10.1016/j.tree.2007.07.009},
  DOI = {10.1016/j.tree.2007.07.009},
  number = {12},
  journal = {Trends in Ecology \& Evolution},
  publisher = {Elsevier BV},
  author = {Rankin,  Daniel J. and Bargum,  Katja and Kokko,  Hanna},
  year = {2007},
  month = dec,
  pages = {643–651}
}

@book{okasha_MLSbook_2006,
  title = {Evolution and the Levels of Selection},
  ISBN = {9780199267972},
  DOI = {10.1093/acprof:oso/9780199267972.001.0001},
  publisher = {Oxford University Press},
  author = {Okasha,  Samir},
  year = {2006},
  month = nov 
}

@article{czuppon_understanding_2021,
	title = {Understanding evolutionary and ecological dynamics using a continuum limit},
	volume = {11},
	issn = {2045-7758},
	doi = {10.1002/ece3.7205},
	number = {11},
	journal = {Ecology and Evolution},
	author = {Czuppon, Peter and Traulsen, Arne},
	year = {2021},
	pages = {5857--5873},
}

@book{gardiner_stochastic_2009,
	address = {Berlin},
	title = {Stochastic methods: a handbook for the natural and social sciences},
	isbn = {978-3-540-70712-7},
	shorttitle = {Stochastic methods},
	publisher = {Springer},
	author = {Gardiner, C. W},
	year = {2009},
}

@article{week_white_2021,
  author   = {Week, Bob and Nuismer, Scott L. and Harmon, Luke J. and Krone, Stephen M.},
  title    = {A white noise approach to evolutionary ecology},
  doi      = {10.1016/j.jtbi.2021.110660},
  issn     = {0022-5193},
  pages    = {110660},
  volume   = {521},
  journal  = {Journal of Theoretical Biology},
  month    = jul,
  year     = {2021},
 }

@article{pawula_approximation_1967,
  author   = {Pawula, R. F.},
  title    = {Approximation of the {Linear} {Boltzmann} {Equation} by the {Fokker}-{Planck} {Equation}},
  doi      = {10.1103/PhysRev.162.186},
  number   = {1},
  pages    = {186--188},
  volume   = {162},
  journal  = {Physical Review},
  year     = {1967},
}

@book{ethier_1986_markov,
  title = {Markov Processes: Characterization and Convergence},
  ISBN = {9780470316658},
  ISSN = {1940-6347},
  url = {http://dx.doi.org/10.1002/9780470316658},
  DOI = {10.1002/9780470316658},
  journal = {Wiley Series in Probability and Statistics},
  publisher = {Wiley},
  author = {Ethier,  Stewart N. and Kurtz,  Thomas G.},
  year = {1986},
}

@article{yagoobi_update_2023,
  title = {Categorizing update mechanisms for graph-structured metapopulations},
  volume = {20},
  ISSN = {1742-5662},
  url = {http://dx.doi.org/10.1098/rsif.2022.0769},
  DOI = {10.1098/rsif.2022.0769},
  number = {200},
  journal = {Journal of The Royal Society Interface},
  author = {Yagoobi,  Sedigheh and Sharma,  Nikhil and Traulsen,  Arne},
  year = {2023},
}

@article{fontanari_groups_2014,
  title = {Nonlinear group survival in Kimura’s model for the evolution of altruism},
  volume = {249},
  ISSN = {0025-5564},
  url = {http://dx.doi.org/10.1016/j.mbs.2014.01.003},
  DOI = {10.1016/j.mbs.2014.01.003},
  journal = {Mathematical Biosciences},
  author = {Fontanari,  José F. and Serva,  Maurizio},
  year = {2014},
  pages = {18–26}
}

@article{vanveelen_groups_2014,
  title = {A simple model of group selection that cannot be analyzed with inclusive fitness},
  volume = {360},
  ISSN = {0022-5193},
  url = {http://dx.doi.org/10.1016/j.jtbi.2014.07.004},
  DOI = {10.1016/j.jtbi.2014.07.004},
  journal = {Journal of Theoretical Biology},
  author = {van Veelen,  Matthijs and Luo,  Shishi and Simon,  Burton},
  year = {2014},
  pages = {279–289}
}

@article{cooney_2022_chromosomes,
  title = {A PDE Model for Protocell Evolution and the Origin of Chromosomes via Multilevel Selection},
  volume = {84},
  ISSN = {1522-9602},
  url = {http://dx.doi.org/10.1007/s11538-022-01062-y},
  DOI = {10.1007/s11538-022-01062-y},
  number = {10},
  journal = {Bulletin of Mathematical Biology},
  publisher = {Springer Science and Business Media LLC},
  author = {Cooney,  Daniel B. and Rossine,  Fernando W. and Morris,  Dylan H. and Levin,  Simon A.},
  year = {2022},
  month = aug 
}

@article{cooney_replicator_2019,
	title = {The replicator dynamics for multilevel selection in evolutionary games},
	volume = {79},
	issn = {1432-1416},
	doi = {10.1007/s00285-019-01352-5},
    number = {1},
	journal = {Journal of Mathematical Biology},
	author = {Cooney, Daniel B.},
	year = {2019},
	pages = {101--154},
}

@article{cooney_long-time_2022,
	title = {Long-time behavior of a {PDE} replicator equation for multilevel selection in group-structured populations},
	volume = {85},
	issn = {1432-1416},
	doi = {10.1007/s00285-022-01776-6},
	number = {2},
	journal = {Journal of Mathematical Biology},
	author = {Cooney, Daniel B. and Mori, Yoichiro},
	year = {2022},
	pages = {12},
}

@article{cooney_evolutionary_2023,
	title = {Evolutionary dynamics within and among competing groups},
	volume = {120},
	doi = {10.1073/pnas.2216186120},
	number = {20},
	journal = {Proceedings of the National Academy of Sciences},
	author = {Cooney, Daniel B. and Levin, Simon A. and Mori, Yoichiro and Plotkin, Joshua B.},
	year = {2023},
	pages = {e2216186120},
}

@article{luo_unifying_2014,
	title = {A unifying framework reveals key properties of multilevel selection},
	volume = {341},
	issn = {0022-5193},
	url = {https://www.sciencedirect.com/science/article/pii/S0022519313004542},
	doi = {10.1016/j.jtbi.2013.09.024},
	journal = {Journal of Theoretical Biology},
	author = {Luo, Shishi},
	year = {2014},
	pages = {41--52},
}

@article{luo_scaling_2017,
	title = {Scaling limits of a model for selection at two scales},
	volume = {30},
	issn = {0951-7715},
	url = {https://doi.org/10.1088/1361-6544/aa5499},
	doi = {10.1088/1361-6544/aa5499},
	number = {4},
	urldate = {2025-12-02},
	journal = {Nonlinearity},
	author = {Luo, Shishi and Mattingly, Jonathan C},
	month = mar,
	year = {2017},
	note = {Publisher: IOP Publishing},
	pages = {1682},
}

@article{boorman_group_1973,
	title = {Group selection on the boundary of a stable population},
	volume = {4},
	issn = {0040-5809},
	doi = {10.1016/0040-5809(73)90007-5},
	number = {1},
	journal = {Theoretical Population Biology},
	author = {Boorman, Scott A. and Levitt, Paul R.},
	year = {1973},
	pages = {85--128},
}

@article{kimura_group_1983,
  title = {Diffusion model of intergroup selection,  with special reference to evolution of an altruistic character},
  volume = {80},
  ISSN = {1091-6490},
  url = {http://dx.doi.org/10.1073/pnas.80.20.6317},
  DOI = {10.1073/pnas.80.20.6317},
  number = {20},
  journal = {Proceedings of the National Academy of Sciences},
  publisher = {Proceedings of the National Academy of Sciences},
  author = {Kimura,  Motoo},
  year = {1983},
  month = Oct,
  pages = {6317–6321}
}

@article{kimura_evolution_1984,
	title = {Evolution of an {Altruistic} {Trait} through {Group} {Selection} as {Studied} by the {Diffusion} {Equation} {Method}},
	volume = {1},
	issn = {0265-0746},
	doi = {10.1093/imammb/1.1.1},
	number = {1},
	journal = {IMA Journal of Mathematics Applied in Medicine \& Biology},
	author = {Kimura, Motoo},
	year = {1984},
	pages = {1--15},
}

@article{yashin_mortality_1985,
	title = {Mortality and aging in a heterogeneous population: {A} stochastic process model with observed and unobserved variables},
	volume = {27},
	issn = {0040-5809},
	shorttitle = {Mortality and aging in a heterogeneous population},
	doi = {10.1016/0040-5809(85)90008-5},
	number = {2},
	journal = {Theoretical Population Biology},
	author = {Yashin, Anatoli I. and Manton, Kenneth G. and Vaupel, James W.},
	month = apr,
	year = {1985},
	pages = {154--175},
}

@article{woodbury_random-walk_1977,
	title = {A random-walk model of human mortality and aging},
	volume = {11},
	issn = {0040-5809},
	doi = {10.1016/0040-5809(77)90005-3},
	number = {1},
	journal = {Theoretical Population Biology},
	author = {Woodbury, Max A. and Manton, Kenneth G.},
	month = feb,
	year = {1977},
	pages = {37--48},
}

@article{woodbury_mathematical_1983,
	title = {A {Mathematical} {Model} of the {Physiological} {Dynamics} of {Aging} and {Correlated} {Mortality} {Selection}. {I}. {Theoretical} {Development} and {Critiques}},
	volume = {38},
	issn = {0022-1422},
	doi = {10.1093/geronj/38.4.398},
	number = {4},
	journal = {Journal of Gerontology},
	author = {Woodbury, Max A. and Manton, Kenneth G.},
	month = jul,
	year = {1983},
	pages = {398--405},
}

@book{gavrilov_biology_1991,
	title = {The Biology of Life Span: A Quantitative Approach},
	isbn = {978-3-7186-4983-9},
	shorttitle = {The biology of life span},
	publisher = {Harwood Academic Publishers},
	author = {Gavrilov, L. A. and Gavrilova, N. S.},
	year = {1991},
}

@article{de_vries_extrinsic_2023,
	title = {Extrinsic mortality and senescence: a guide for the perplexed},
	volume = {3},
	issn = {2804-3871},
	shorttitle = {Extrinsic mortality and senescence},
	url = {https://peercommunityjournal.org/articles/10.24072/pcjournal.253/},
	doi = {10.24072/pcjournal.253},
	urldate = {2024-01-27},
	journal = {Peer Community Journal},
	author = {de Vries, Charlotte and Galipaud, Matthias and Kokko, Hanna},
	year = {2023},
}

@article{nielsen_gompertz_2024,
	title = {The {Gompertz} {Law} emerges naturally from the inter-dependencies between sub-components in complex organisms},
	volume = {14},
	copyright = {2024 The Author(s)},
	issn = {2045-2322},
	url = {https://www.nature.com/articles/s41598-024-51669-5},
	doi = {10.1038/s41598-024-51669-5},
	
	number = {1},
	urldate = {2024-07-01},
	journal = {Scientific Reports},
	author = {Nielsen, Pernille Yde and Jensen, Majken K. and Mitarai, Namiko and Bhatt, Samir},
	month = jan,
	year = {2024},
	note = {Publisher: Nature Publishing Group},
	pages = {1196},
}

@article{ledberg_exponential_2020,
	title = {Exponential increase in mortality with age is a generic property of a simple model system of damage accumulation and death},
	volume = {15},
	issn = {1932-6203},
	url = {https://journals.plos.org/plosone/article?id=10.1371/journal.pone.0233384},
	doi = {10.1371/journal.pone.0233384},
	
	number = {6},
	urldate = {2024-08-21},
	journal = {PLOS ONE},
	author = {Ledberg, Anders},
	month = jun,
	year = {2020},
	note = {Publisher: Public Library of Science},
	pages = {e0233384},
}

@article{le_bras_lois_1976,
	title = {Lois de mortalité et age limite},
	volume = {31},
	issn = {0032-4663},
	url = {https://www.jstor.org/stable/1530761},
	doi = {10.2307/1530761},
	number = {3},
	urldate = {2024-09-03},
	journal = {Population (French Edition)},
	author = {Le Bras, Hervé},
	year = {1976},
	note = {Publisher: Institut National d'Études Démographiques},
	pages = {655--692},
}

@article{gavrilov_reliability_2001,
	title = {The {Reliability} {Theory} of {Aging} and {Longevity}},
	volume = {213},
	issn = {0022-5193},
	url = {https://www.sciencedirect.com/science/article/pii/S0022519301924300},
	doi = {10.1006/jtbi.2001.2430},
	number = {4},
	urldate = {2024-09-03},
	journal = {Journal of Theoretical Biology},
	author = {Gavrilov, L. A. and Gavrilova, N. S.},
	month = dec,
	year = {2001},
	pages = {527--545},
}

@article{bhat_stochastic_2025,
	title = {A stochastic field theory for the evolution of quantitative traits in finite populations},
	volume = {161},
	doi = {10.1016/j.tpb.2024.10.003},
	journal = {Theoretical Population Biology},
	author = {Bhat, Ananda Shikhara},
	year = {2025},
	pages = {1--12},
}
\end{refcontext}
\end{refsection}

\end{document}